%% file: epi0mupi0.tex
\documentclass[aps,prd,twocolumn,superscriptaddress,showpacs,floatfix,nofootinbib]{revtex4-1}

\usepackage{graphicx}
\usepackage{epsfig}
\usepackage{subfigure}
\usepackage{xspace}
\usepackage{dcolumn}
\usepackage{multirow}
\usepackage{longtable}
\usepackage{amsmath}
\usepackage{mathtools}
\usepackage{xcolor}
\usepackage{lineno}
\usepackage{grffile}
\usepackage{hyperref}
\hypersetup{
	colorlinks=true, 
	pdfborder={0 0 0},
	linkcolor=blue,
	citecolor=blue,
	urlcolor=blue,
}

\begin{document}

\title{Search for proton decay via $p\rightarrow e^+\pi^0$ and $p\rightarrow \mu^+\pi^0$ with an enlarged fiducial volume in Super-Kamiokande I-IV}

\input{sk_authors_epi0mupi02019.tex}

\date{\today}

\begin{abstract}
We have searched for proton decay via $p\rightarrow e^+\pi^0$ and $p\rightarrow \mu^+\pi^0$ modes with the enlarged fiducial volume data of Super-Kamiokande from April 1996 to May 2018, which corresponds to 450~kton$\cdot$years exposure.
We have accumulated about 25\% more livetime and enlarged the fiducial volume of the Super-Kamiokande detector from 22.5~kton to 27.2~kton for this analysis, so that 144~kton$\cdot$years of data, including 78~kton$\cdot$years of additional fiducial volume data, has been newly analyzed.
No candidates have been found for $p\rightarrow e^+\pi^0$ and one candidate remains for $p\rightarrow \mu^+\pi^0$ in the conventional 22.5~kton fiducial volume and it is consistent with the atmospheric neutrino background prediction.
We set lower limits on the partial lifetime for each of these modes: $\tau/B(p\rightarrow e^+\pi^0) > 2.4 \times 10^{34}$~years and $\tau/B(p\rightarrow \mu^+\pi^0) > 1.6 \times 10^{34}$~years at 90\% confidence level.
\end{abstract}

\maketitle


\section{Introduction}
Grand Unified Theories (GUTs)~\cite{su5} extend the Standard Model gauge symmetry to larger 
symmetry groups and provide explanations for the quantization of electric charge and predict the convergence 
of the electromagnetic, weak, and strong interaction couplings at energies around $\sim10^{16}$~GeV~\cite{coupling}.
There are a variety of proposed GUT models based on different gauge groups, such as SU(5)~\cite{su5}, SO(10)~\cite{so10}, E$_6$~\cite{e6}, some with and some without incorporating supersymmetry. 
These models offer dark matter candidates~\cite{dm}, neutrino mass generation mechanisms~\cite{numass}, 
and insight into the strong CP problem~\cite{axion}, making them a promising target for many searches for physics beyond the Standard Model.    
Furthermore, by incorporating quarks and leptons into common multiplets, GUTs generically predict transitions between the two which result in baryon-number-violating proton decays, whose observation would provide strong support for such models.

While the $p\rightarrow e^+\pi^0$ mode is favored by many GUT models, the ``flipped SU(5)'' ones~\cite{flip}
predict a higher branching fraction in the $p\rightarrow \mu^+\pi^0$ channel. 
Since recent theoretical studies~\cite{muramatsu,nagata} show that the preferred models of GUTs may be revealed by the first signs of proton decay, it is important to search for both modes.
Both modes produce back-to-back event topologies in which all final state particles are visible 
making it possible to cleanly separate a proton decay signal from atmospheric neutrino backgrounds in 
water Cherenkov detectors. 
Moreover, free protons (hydrogen nuclei) in their water provide enhanced background rejection capabilities 
since a decay therein would be free from the effects of the Fermi motion and intranuclear scattering processes that alter the final 
state particles from decays within $^{16}$O. 

These decay modes have been the target of several experimental searches, but there have been no 
positive observations so far~\cite{past}.
Prior to the present work the most stringent constraints come from 306~kton$\cdot$years of Super-Kamiokande data: 
$\tau/B(p\rightarrow e^+\pi^0) > 1.6 \times 10^{34}$~years and $\tau/B(p\rightarrow \mu^+\pi^0) > 7.7 \times 10^{33}$~years at 90\% confidence level~\cite{last}.
Since then the detector has accumulated about 25\% more live days of data and its 
fiducial volume has been enlarged from 22.5~kton to 27.2~kton resulting in a roughly 50\% larger exposure of 450~kton$\cdot$years. 
This paper describes a search for these two proton decay modes using this updated data set 
and is organized as follows.
A summary of the Super-Kamiokande detector and its event reconstruction is presented in Section~\ref{second} 
before describing analysis improvements that enabled the fiducial volume to be enlarged in Section~\ref{enlarge}.
Section~\ref{simu} details the simulation of proton decay and atmospheric neutrino events 
and Sections~\ref{perform} and~\ref{results} describe the proton decay search sensitivity and results.
As no proton decay signal has been found, lifetime limits are calculated in Section~\ref{calc} 
before concluding in Section~\ref{conclude}.

\section{The Super-Kamiokande Detector}
\label{second}
Super-Kamiokande (SK) is located about 1,000~m (2,700 meters water equivalent) under Mt. Ikenoyama in, Gifu Prefecture, Japan. 
The detector is an upright cylindrical vessel, 39.3~m in diameter and 41.4~m in height, and is filled 
with 50~kton of ultrapure water that is optically separated into two regions, an inner detector (ID) and an outer detector (OD). 
The ID has a diameter of 33.8~m and a height of 36.2~m that is viewed
by more than 11,000 inward-facing 50-cm photomultiplier tubes (PMTs) mounted on a steel structure offset from the tank outer wall by 2~m and forming the boundary with the OD.

The OD is composed of 20-cm PMTs installed behind the ID PMTs and facing outward. 
Reflective Tyvek lines the walls of the OD and wavelength shifting plates are mounted on its PMTs in order 
to achieve more efficient light collection and improve the rejection of external backgrounds. 
Detailed descriptions of the detector and its calibration can be found in~\cite{skdetector, skcalib}. 

The SK data are divided into four phases representing different configurations of the detector, \mbox{SK-I}, \mbox{-II}, \mbox{-III}, and \mbox{-IV}.
The \mbox{SK-I} period began in April 1996 and ended in July 2001 with a photocathode coverage of 40\%. 
In November 2001, a chain reaction implosion destroyed more than half of the PMTs,
such that \mbox{SK-II} was operated from October 2002 to October 2005, using 5,182 ID PMTs with 19\% photocathode coverage.
Since \mbox{SK-II} the ID PMTs have been covered in fiber-reinforced plastic cases with an acrylic window to prevent similar accidents.
New ID PMTs were installed thereafter and \mbox{SK-III} started in July 2006 with 40\% of photocathode coverage.
In September 2008 the front-end electronics~\cite{qbee} and data acquisition system were upgraded to start 
the \mbox{SK-IV} period, which continued until May 2018.
These systems improve the efficiency for tagging Michel electrons and the 
faint 2.2~MeV gamma ray that accompanies neutron capture on hydrogen, 
resulting in 20\% higher signal selection efficiency in the search for $p\rightarrow \mu^+\pi^0$ decays 
and a reduction of atmospheric neutrino backgrounds in both decay modes considered here by 50\% 
relative to the previous analysis~\cite{last}.

Charged particles above the Cherenkov threshold inside the ID are reconstructed using PMT timing and charge information.
First, a charged particle's initial vertex is reconstructed by finding the point 
for which the distribution of time-of-flight corrected PMT times has the sharpest peak.
This is done by maximizing the estimator, 
\begin{equation}
\label{goodness}
{\rm goodness} = \sum_{i}\frac{1}{\sigma^{2}_{i}(q_{i})}\exp{\left(-\frac{(t^{\prime}_{i}-t_{0})^{2}}{2\times(\langle\sigma\rangle\times1.5)^{2}}\right)},
\end{equation}
\noindent where $\sigma_{i}$ is the timing resolution of the $i$th PMT tabulated as a function of observed charge,
 $\langle\sigma\rangle$ is the average timing resolution for hit PMTs, 
and $t^{\prime}_i$ is the $i$th PMT's residual time for an assumed vertex position.
In the residual time calculation, the track length of the charged particle is taken into account only for the most energetic one since this is prior to the Cherenkov ring counting algorithm described below.
Here 1.5 is a tuning parameter that has been chosen to optimize performance 
and $t_0$ is chosen so that the goodness is maximal at each tested vertex position.
The vertex position is the point with the highest goodness.

After reconstructing the vertex position, 
the number of Cherenkov rings projected on the ID wall is determined using a ring pattern recognition algorithm based on 
the Hough transformation~\cite{hough}.
The PMT charge distribution is corrected for the water's attenuation length and PMT acceptance before 
being  transformed into the Hough space. 
Ring centers, corresponding to particle directions, and ring opening angles, corresponding to Cherenkov angles, 
manifest as peaks in the resulting distribution. 

Each identified Cherenkov ring is identified as either showering ($e$-like) particle ($e^{\pm}, \gamma$) or non-showering ($\mu$-like) particle ($\mu^{\pm}, \pi^{\pm}$) based on the pattern of hit PMTs in the ring. 
Rings from electrons and gamma rays tend to have diffuse edges due to 
the overlap of many Cherenkov rings produced by the particles in their electromagnetic showers.
On the other hand, muon and pion rings tend to have crisp edges since they do not shower due to their larger masses. 
The expected charge distribution is calculated for each particle assumption and compared to the observation 
using 
\begin{equation}
\label{pid}
\chi^{2} \propto -\sum_{\theta_i<(1.5\times\theta_c)}\log_{10} P(q_{i},q^{\rm exp}_{i}),
\end{equation}
\noindent 
where $q_i$ is the $i$th PMT's observed charge, $q^{\rm exp}_{i}$ is the expected charge for that PMT for either the electron or muon assumption, and $P$ is the probability of observing $q_i$ given an expectation of $q^{\rm exp}_{i}$.
For both the electron and muon assumptions, the summation is performed for PMTs 
whose angle to the ring direction ($\theta_{i}$) is within 1.5 times the reconstructed Cherenkov opening angle ($\theta_{c}$).
The particle type with the smallest $\chi^{2}$ is assigned to each reconstructed ring.
For multi-ring events such as those from $p\rightarrow e^+\pi^0$ and $p\rightarrow \mu^+\pi^0$ decays, 
the observed charge at each PMT is separated into the contribution from each Cherenkov ring using the expected charge distribution.
In this step, the contributions from light scattering in the water and reflection on the PMT surfaces are calculated 
and subtracted from the total charge associated with each ring in order to estimate their momenta.
The relationship between the total observed charge within a 70 degree half opening angle from the particle direction and particle momentum is based on Monte Carlo (MC) 
simulation and is tabulated in advance for each particle 
taking into account corrections for water attenuation length and PMT acceptance.
A more detailed description of the event reconstruction algorithm can be found in~\cite{recon}.
The performance of the reconstruction on proton decay events is discussed in the next section. 

Neutrons emitted from a primary event thermalize in water and are eventually captured by hydrogen,
resulting in the emission of a 2.2~MeV gamma ray.
Such gamma rays are tagged by searching for clusters of at least five hit ID PMTs 
in a sliding 10~nsec window scanned 
between 18~$\mu$sec and 535~$\mu$sec after the primary event trigger.
This search time region is set to avoid effects from PMT after-pulsing which occurs between 12 and 18~$\mu$sec after the primary PMT hit.
A dedicated neural network that has been trained using simulated atmospheric neutrino events and random trigger data representing mostly PMT dark noise is applied to each cluster to identify it as either a neutron signal or background. 
The total neutron tagging efficiency is estimated to be $25.2 \pm 2.3\%$ using atmospheric neutrino MC  
and calibration data from an AmBe neutron source. 
The estimated false-positive rate of the algorithm is 0.018~neutron-candidates per primary event. 
A detailed description of the neural net and recent update of the algorithm can be found in~\cite{ntag,matanaka}.

\section{Enlarging The Fiducial Volume}
\label{enlarge}
This analysis uses events termed ``fully contained'' (FC) whose interaction vertex has been reconstructed within the 
ID and which have no cluster of hits in the OD and which pass other selection criteria~\cite{sk1full}. 
The detector's fiducial volume is defined using the distance between the nearest ID wall and the event's 
reconstructed vertex ($dwall$) without reconstructed particle direction dependence. 
In the previous analysis~\cite{last} this region was defined as the region more than 200~cm from the wall (200~cm $< dwall$) which corresponds to 22.5~kton of fiducial mass.
In order to further improve the proton decay search sensitivity, the following studies have been conducted for the present analysis 
to enlarge the fiducial volume boundary from 200~cm to 100~cm, thereby increasing the water mass to 27.2~kton.
In the following, the term ``conventional'' fiducial volume refers to the region with $dwall$ greater than 200~cm 
and ``additional'' fiducial volume refers to the region with $dwall$ between 100~cm and 200~cm. 
Terms used to represent different detector regions are summarized in Table~\ref{tab:term}.

\begin{table}[htbp]
\caption{Terms used to represent detector regions.}
\begin{tabular}{lcc}
\hline
Term & $dwall$ & water mass \\
\hline
Conventional & 200~cm $< dwall$ & 22.5~kton \\
Additional & 100 $<dwall\leq$ 200~cm & 4.7~kton \\
Enlarged & 100~cm $<dwall$ & 27.2~kton \\
Outside & 50 $<dwall\leq$ 100~cm & 2.6~kton\\
\hline
\end{tabular}
\label{tab:term}
\end{table}

Non-neutrino background events originating from outside of the conventional fiducial volume, 
such as cosmic-ray muons or noise events from erroneous discharges in a PMT's dynode (``flasher'' events), 
typically have reconstructed vertices on the wall ($dwall$ around 0~cm) 
but due to the limited vertex resolution, some may be reconstructed within the fiducial volume. 
All FC events with reconstructed vertices more than 50~cm from the wall 
were eye-scanned using a graphical event display tool to estimate the contamination from such backgrounds.
Since the enlarged fiducial volume boundary is closer to the wall, these backgrounds occur more frequently and have been identified 
by this scanning. 
No event has been rejected based on the scanning results but
several changes have been made to the existing FC event selection~\cite{sk1full} to remove them. 

First, to reject cosmic-ray muons that pass though cable bundles running through the OD without leaving sufficient light to trigger its PMTs before entering the ID, plastic scintillator-based veto counters are placed above the four bundles located closest to the ID.
In previous analyses, events with both a signal in a veto counter 
and a vertex less than 4~m from the top of the cable bundle have been rejected. 
In order to more accurately identify such entering particles, an independent algorithm 
dedicated to muon vertex reconstruction that forces the vertex (entering point) to be reconstructed on the ID wall is used.
In the present work, events leaving sufficiently large charge in a veto counter are also removed regardless of their vertex position.
The veto counters were installed in the middle of \mbox{SK-I}, Apr. 1997 and the cut associated with them is applied to the data since then.
In addition, events whose reconstructed direction is downward-going ($\cos\theta_{\rm{z}}>0.6$, where $\cos\theta_{z}=1$ is vertically downward-going) and whose vertex is less than 2.5~m from any of the 12 ports housing a cable bundle are removed.
Here again the dedicated vertex algorithm is used.
The above two changes eliminate cosmic-ray muons passing through the cable bundles. \par

Events are also removed if their vertex goodness as defined in Equation~(\ref{goodness}) is less than 0.77 and if they have 
more than 7000 ID PMT hits with more than 70000~photoelectrons deposited therein (corresponding to about 7~GeV of energy deposition in the detector) and more than 5 OD PMT hits around their entrance point.
This cut eliminates high energy cosmic-ray muons that have been mis-reconstructed inside the fiducial volume.
In order to reject Michel electrons from cosmic-ray muons which are below the Cherenkov threshold and stop in the ID, 
events which have more than 50 (55 in \mbox{SK-IV}) OD hits in a sliding 200~nsec timing window before an ID trigger are removed.

\begin{figure}[htbp]
\centering
\includegraphics[width=0.5\textwidth, clip, viewport =  0.00   0.00   567.00   384.00]{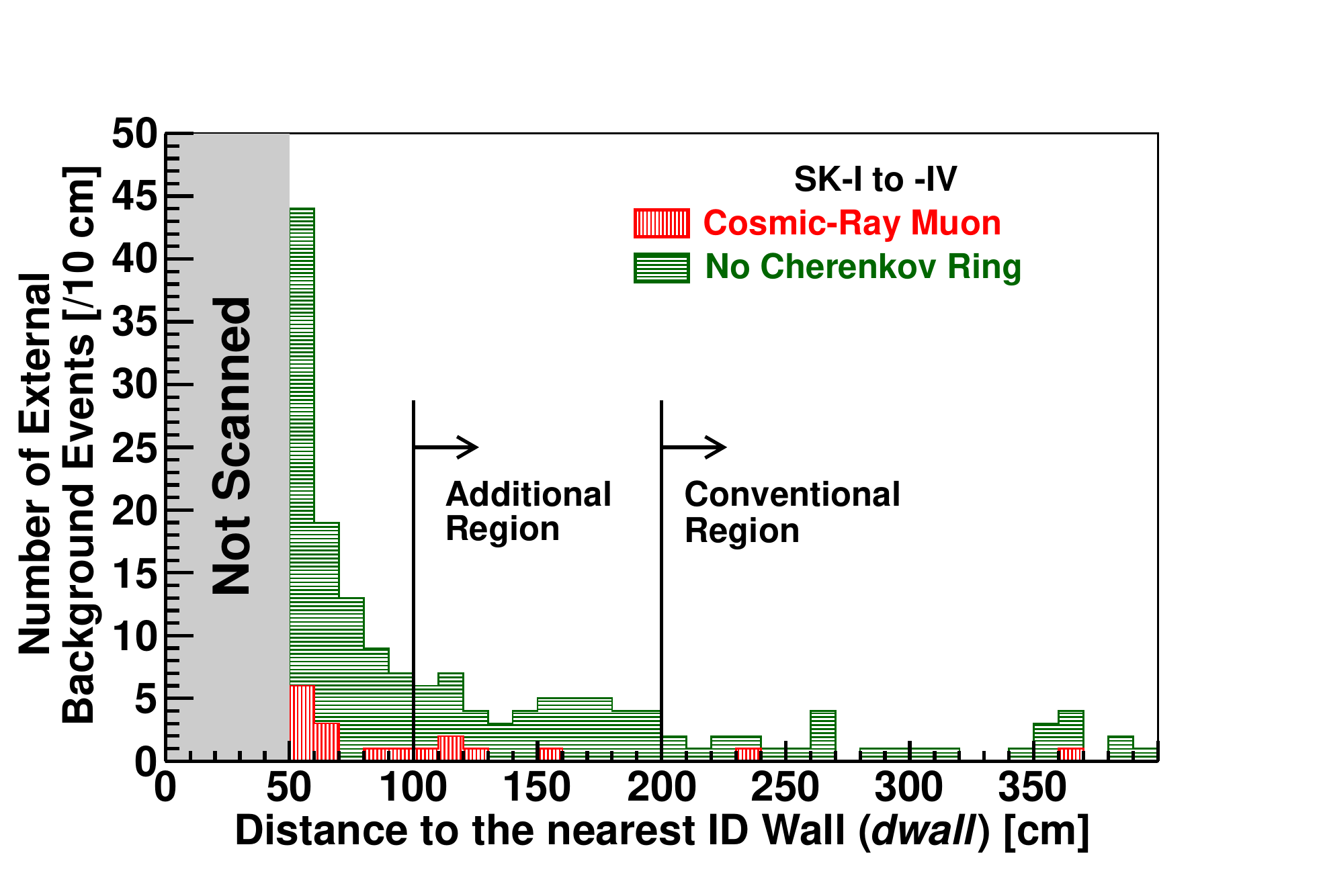}
\caption{Distribution of the distance to the nearest ID wall from the reconstructed vertex ($dwall$) of 
non-neutrino background events remaining in \mbox{SK-I} to \mbox{-IV}. 
Cosmic-ray muon events (red, vertical stripe) and no Cherenkov ring events (green, horizontal stripe) are shown in a stacked histogram.
No Cherenkov ring events are dominated by PMT ``flasher'' events described in the text.
}
\label{fig:nonnubg}
\end{figure}

Figure~\ref{fig:nonnubg} shows the remaining non-neutrino backgrounds in the \mbox{SK-I} to \mbox{-IV} data as a function of $dwall$ and identified 
by the visual scanning after adopting these new criteria.
With the new selection the fraction of non-neutrino backgrounds relative to atmospheric neutrino events is about 0.5\% (0.1\%) in the 
additional (conventional) fiducial volume, tolerably worse than that in the conventional fiducial volume.
However, this fraction increases to 2.0\% for $dwall$ between 50~cm and 100~cm from the wall and restricts further expansion of the fiducial volume.
The region with $dwall$ between 50~cm and 100~cm is referred to ``outside'' region in the following.
The new selection has negligible impact on signal efficiency for atmospheric neutrinos and proton decay. 

For events occurring in the region close to the ID wall, the number of hit PMTs is typically small and more localized than for interactions in the conventional fiducial volume, and it is easier for particles to escape the ID with non-negligible momentum.
It is the case for both $p\rightarrow e^+\pi^0$ and $p\rightarrow \mu^+\pi^0$ decays because they produce back-to-back event topologies and one of the reconstructed rings generally has a small distance to the wall along its direction.
As shown in Equations~\ref{goodness} and~\ref{pid}, since the accuracy of the vertex reconstruction and particle identification 
depend on the number of ID PMT hits, both will degrade with fewer PMT hits. 
Furthermore, where the PMT hits are too localized, the reconstruction algorithm may be unable to find or separate 
Cherenkov rings from multiple particles. 
Particles that exit the ID will have biased reconstructed momenta and make it impossible to tag any Michel electrons from their decays.
 
\begin{figure}[htbp]
\centering
\includegraphics[width=0.5\textwidth, clip, viewport =  0.00   0.00   567.00   384.00]{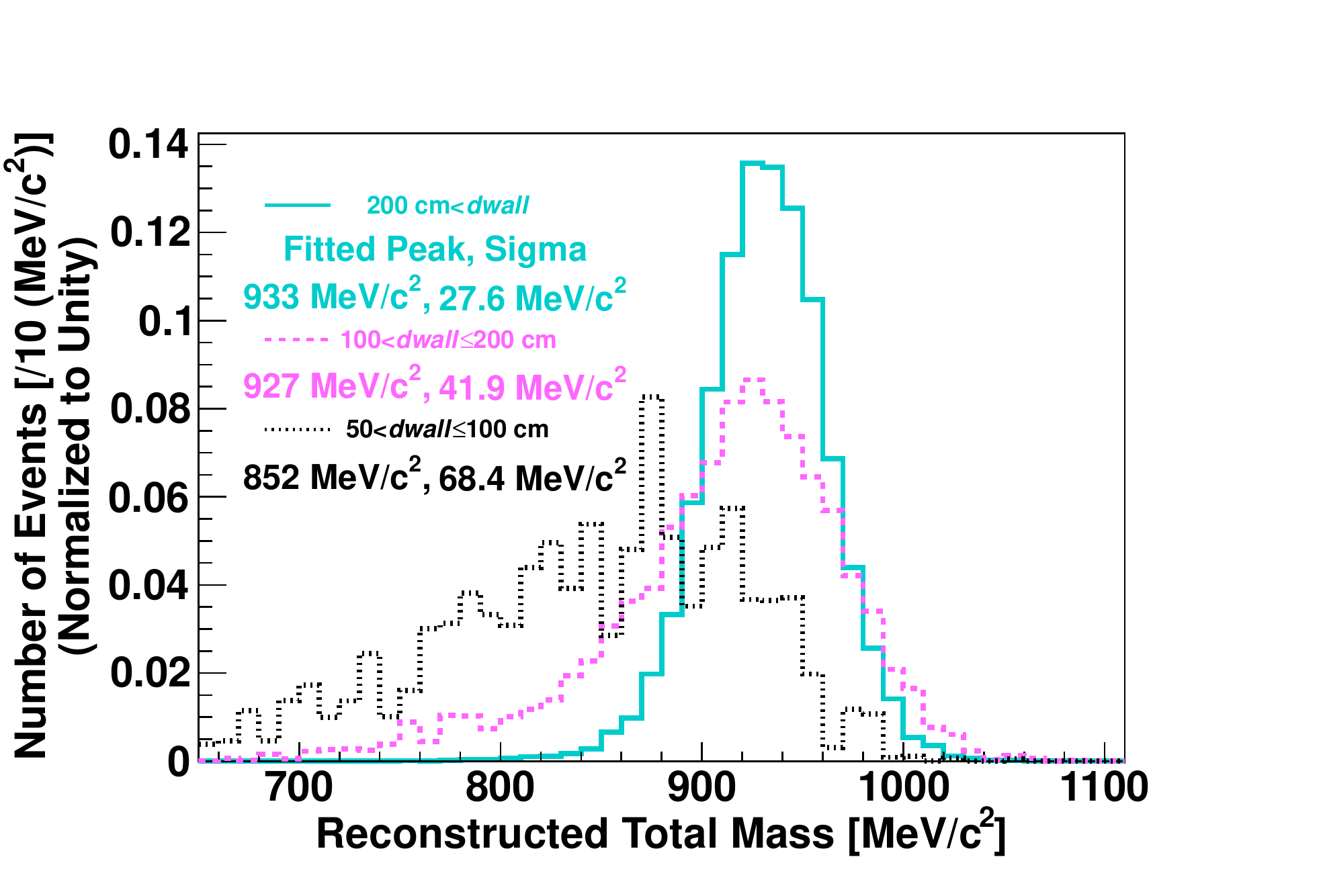}
\caption{Reconstructed total mass distribution for free proton decays via $p\rightarrow e^+\pi^0$ MC events with three rings.
The solid cyan histogram shows events reconstructed in the conventional fiducial volume (200~cm $< dwall$), 
the pink dashed histogram shows that for the additional fiducial volume (100 $< dwall \leq$ 200~cm) and 
the black dotted histogram is for the region between $dwall$ of 50~cm and 100~cm. 
Gaussian fits have been used to determine the mass peak and width.
}
\label{fig:epi0tmass3}
\end{figure}

Though these issues are unavoidable, their impact has been mitigated by the following improvements to the particle identification calculation.
With a lower number of ID PMT hits, the expected charge calculation ($q^{\rm exp}_i$ in Equation~\ref{pid}) for each PMT becomes more important.
Therefore the expected charge tables have been updated using events from both within and outside of the conventional 
fiducial volume and have been parameterized as a function of the distance between the event vertex and PMT position 
as well as the angle to the particle direction as viewed by the PMT.
By adopting the updated charge tables, the particle mis-identification probability in the additional fiducial volume 
has been reduced by about 35\% for single-ring events and for proton decay MC events, 
the signal selection efficiency for both modes has been improved by about 20\% relative to previous versions. 
Table~\ref{tab:recon} compares the reconstruction performance among the different ID regions 
using free proton decay ($p\rightarrow e^+\pi^0$ and $p\rightarrow \mu^+\pi^0$) MC events to remove the influence of nuclear effects.
Though the full event selection is presented below, note that two-ring events are included in the 
analysis to allow for asymmetric $\pi^{0}$ decays in which one of the gamma ray rings may overlay with that 
from another particle.
The fraction of these events near the ID wall is larger than that in the conventional fiducial volume due 
to the increased likelihood of missing a third ring resulting from more localized PMT hits.
For the $p\rightarrow \mu^+\pi^0$ mode, an escaping muon causes a lower Michel electron tagging efficiency for events close to the ID wall.
Figure~\ref{fig:epi0tmass3} shows the reconstructed total mass distribution of $p\rightarrow e^+\pi^0$ MC events with 
three reconstructed rings from free proton decays. 
Here all proton decay selection cuts (defined in Sec.~\ref{perform}) except the cut on the plotted variable have been applied. 
The two distributions outside of the conventional fiducial volume have longer tails in the low mass region due to particles exiting the ID.

\begin{table*}[htbp]
\centering
\caption{Summary of reconstruction performance in different ID regions using only free proton decay MC events 
and weighted by the combined \mbox{SK-I} to \mbox{-IV} livetime.
The particle identification (PID) efficiency is the fraction of events which pass the proton decay signal criterion {\bf C3} defined in Sec.~\ref{perform} out of all two- and three-ring signal events.
Here $M_{\pi^{0}}$ ($M_{{\rm tot}}$) peak is the reconstructed neutral pion (total) mass distribution's peak position after 
applying all selections except {\bf C6}.  
The peak is determined using a Gaussian fit.
Note that $M_{\pi^{0}}$ peak values are evaluated for only three-ring events.
The terms ``Conventional'', ``Additional'' and ``Outside'' stand for the conventional (200~cm $< dwall$), additional fiducial volume (100 $< dwall \leq$ 200~cm) and outside region (50 $< dwall \leq$ 100~cm), respectively.
The outside region is not used for the present analysis.
}
\begin{tabular}{lcccccccc} \hline  \hline
Decay Mode \hspace{0mm} & Vertex \hspace{0mm} & 2-ring \hspace{0mm} & 3-ring \hspace{0mm} & PID \hspace{0mm} & Michel electron \hspace{0mm} & $M_{\pi^0}$ peak & $M_{{\rm tot}}$ peak\hspace{0mm} & $M_{{\rm tot}}$ peak\\
Region \hspace{0mm} & resolution \hspace{0mm} & fraction \hspace{0mm} & fraction \hspace{0mm} & efficiency \hspace{0mm} & tagging efficiency \hspace{0mm} & (3-ring) & (2-ring) \hspace{0mm} & (3-ring)\\ \hline
$p\rightarrow e^+\pi^0$  \\
Conventional & 17.1~cm & 39.3\% & 58.7\% & 95.8\% & N.A. & \hspace{1mm}135 ${\rm MeV}/c^{2}$\hspace{1mm} & \hspace{1mm}910 ${\rm MeV}/c^{2}$\hspace{1mm} & \hspace{1mm}933 ${\rm MeV}/c^{2}$\hspace{1mm} \\
Additional & 24.1~cm & 51.9\% & 44.5\% & 87.7\% & N.A. &  \hspace{1mm}134 ${\rm MeV}/c^{2}$\hspace{1mm} & \hspace{1mm}898 ${\rm MeV}/c^{2}$\hspace{1mm} & \hspace{1mm}927 ${\rm MeV}/c^{2}$\hspace{1mm} \\
Outside & 25.9~cm & 55.9\% & 37.6\% & 89.6\% & N.A. & \hspace{1mm}\multirow{1}{*}{125 ${\rm MeV}/c^{2}$}\hspace{1mm} & \hspace{1mm}828 ${\rm MeV}/c^{2}$\hspace{1mm} & \hspace{1mm}852 ${\rm MeV}/c^{2}$ \hspace{1mm}\\ \hline
$p\rightarrow \mu^+\pi^0$  \\
Conventional & 20.6~cm & 36.8\% & 62.0\% & 96.2\% & 88.6\% & \hspace{1mm}135 ${\rm MeV}/c^{2}$\hspace{1mm} & \hspace{1mm}917 ${\rm MeV}/c^{2}$\hspace{1mm} & \hspace{1mm}936 ${\rm MeV}/c^{2}$\hspace{1mm} \\
Additional & 24.8~cm & 49.1\% & 48.1\% & 90.7\% & 78.9\% & \hspace{1mm}132 ${\rm MeV}/c^{2}$\hspace{1mm} & \hspace{1mm}917 ${\rm MeV}/c^{2}$\hspace{1mm} & \hspace{1mm}941 ${\rm MeV}/c^{2}$\hspace{1mm}\\
Outside & 26.7~cm & 55.3\% & 39.8\% & 82.6\% & 64.8\% & \hspace{1mm}\multirow{1}{*}{125 ${\rm MeV}/c^{2}$}\hspace{1mm} & \hspace{1mm}905 ${\rm MeV}/c^{2}$\hspace{1mm} & \hspace{1mm}925 ${\rm MeV}/c^{2}$\hspace{1mm}\\ \hline \hline
\end{tabular}
\label{tab:recon}
\end{table*}

\begin{figure*}[htbp]
\centering
\includegraphics[width=0.8\textwidth, clip, viewport =  0.00   0.00   567.00   466.00]{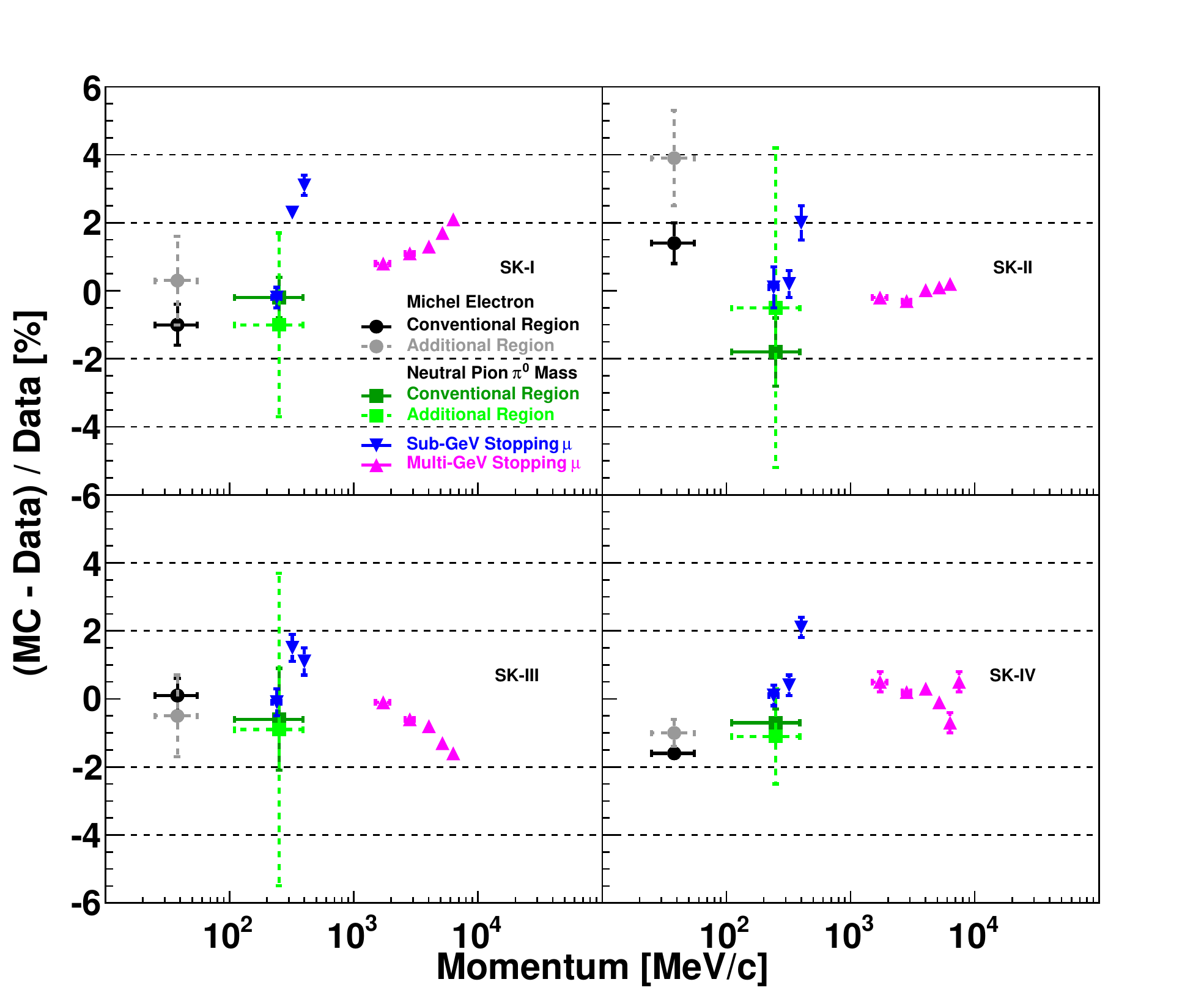}
\caption{Absolute energy scale difference between data and MC control samples.
 Vertical error bars denote the statistical uncertainty and horizontal error bars show 
the momentum range for each control sample. 
For the Michel electron and $\pi^{0}$ samples, points with solid error bars correspond to the conventional fiducial volume and points with dashed error bars correspond to the additional fiducial volume.
Cosmic-ray muon samples are considered for the enlarged fiducial volume.
The term ``Sub-GeV'' (``Multi-GeV'') stands for energy deposition in the detector of less (more) than 1.33~GeV.
}
\label{fig:escale}
\end{figure*}

\begin{figure}[htbp]
\centering
\includegraphics[width=0.5\textwidth, clip, viewport =  0.00   0.00   567.00   557.00]{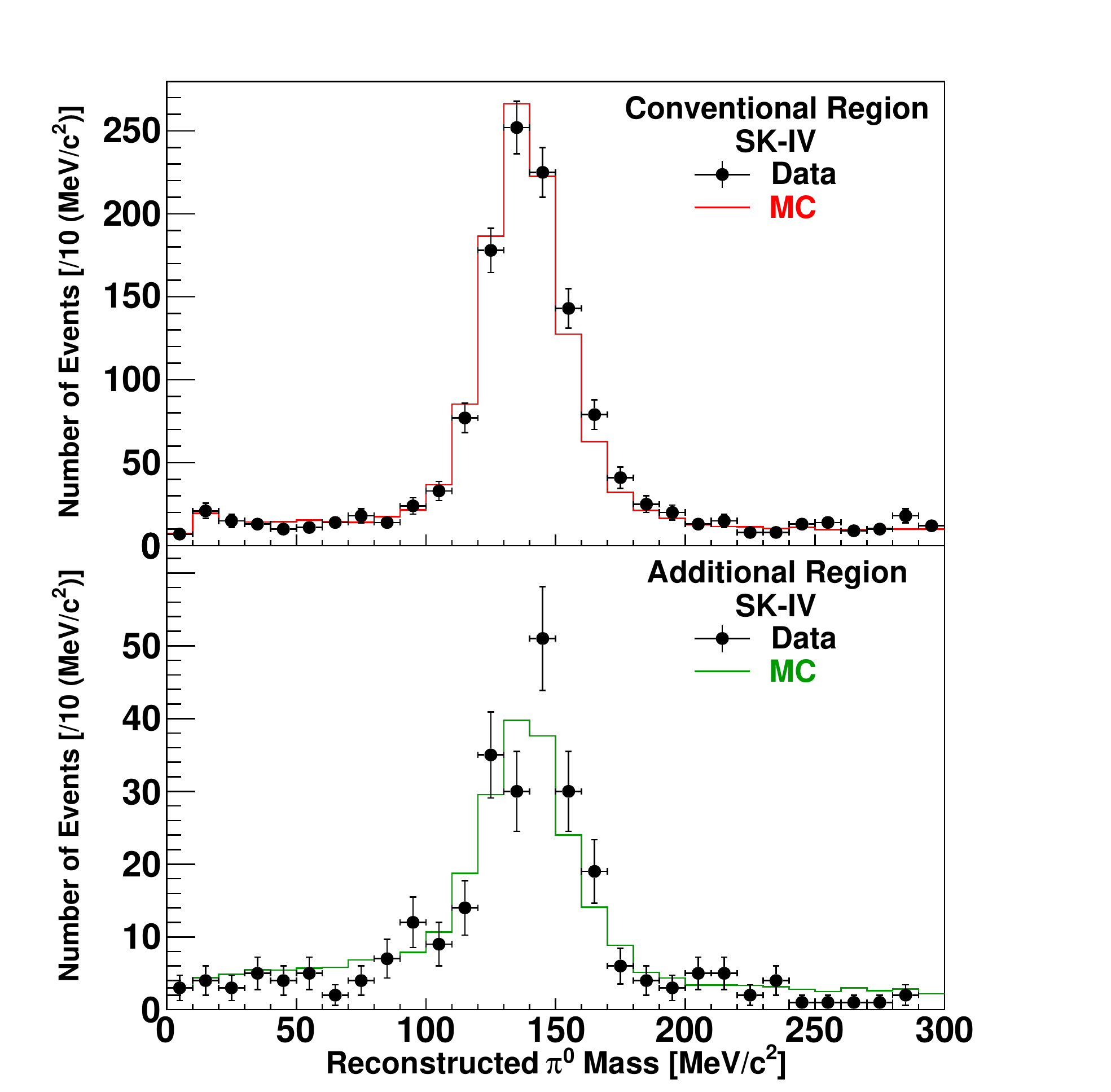}
\caption{
Reconstructed $\pi^{0}$ mass distribution of the two-ring both $e$-like events in \mbox{SK-IV}.
The top plot corresponds to the conventional and the bottom plot corresponds to the additional fiducial volume.
The black dots show the data in \mbox{SK-IV}, and red and green histograms show the atmospheric neutrino MC events.
Vertical error bars denote the statistical uncertainty.
}
\label{fig:pi0mass}
\end{figure}

Systematic uncertainties associated with the reconstruction in the additional fiducial volume are estimated 
separately from those in the conventional fiducial volume.
In this analysis the energy scale uncertainty is the dominant error from the reconstruction and 
it is estimated with the difference in reconstructed momenta or masses between data and MC for several control samples in each SK phase.
For the absolute energy scale, the Michel electron momentum spectrum from stopping cosmic-ray muons,
the $\pi^{0}$ mass spectrum from neutral current atmospheric neutrino interactions, and the momentum divided by the range of 
cosmic-ray muons with energy deposition in the detector of more and less than 1.33~GeV are used. 
Figure~\ref{fig:escale} shows the absolute energy scale difference between data and MC for each of these control samples.
For Michel electrons and $\pi^{0}$ events, the event vertices can be reconstructed inside the ID and 
are used to estimate the difference between data and MC for each volume of the detector separately. 
Figure~\ref{fig:pi0mass} shows the reconstructed $\pi^{0}$ mass distributions for both the conventional and additional fiducial volumes in \mbox{SK-IV}.
Two-ring both $e$-like atmospheric neutrino events are used and they confirm good agreement between data and MC.
The absolute scale uncertainty is taken to be the value of the most discrepant control sample in each SK phase and fiducial volume.
To take the most discrepant control sample, cosmic-ray muon samples are considered for both fiducial volumes.

\begin{figure}[htbp]
\centering
\includegraphics[width=0.5\textwidth, clip, viewport =  00.00   0.00   567.00   384.00]{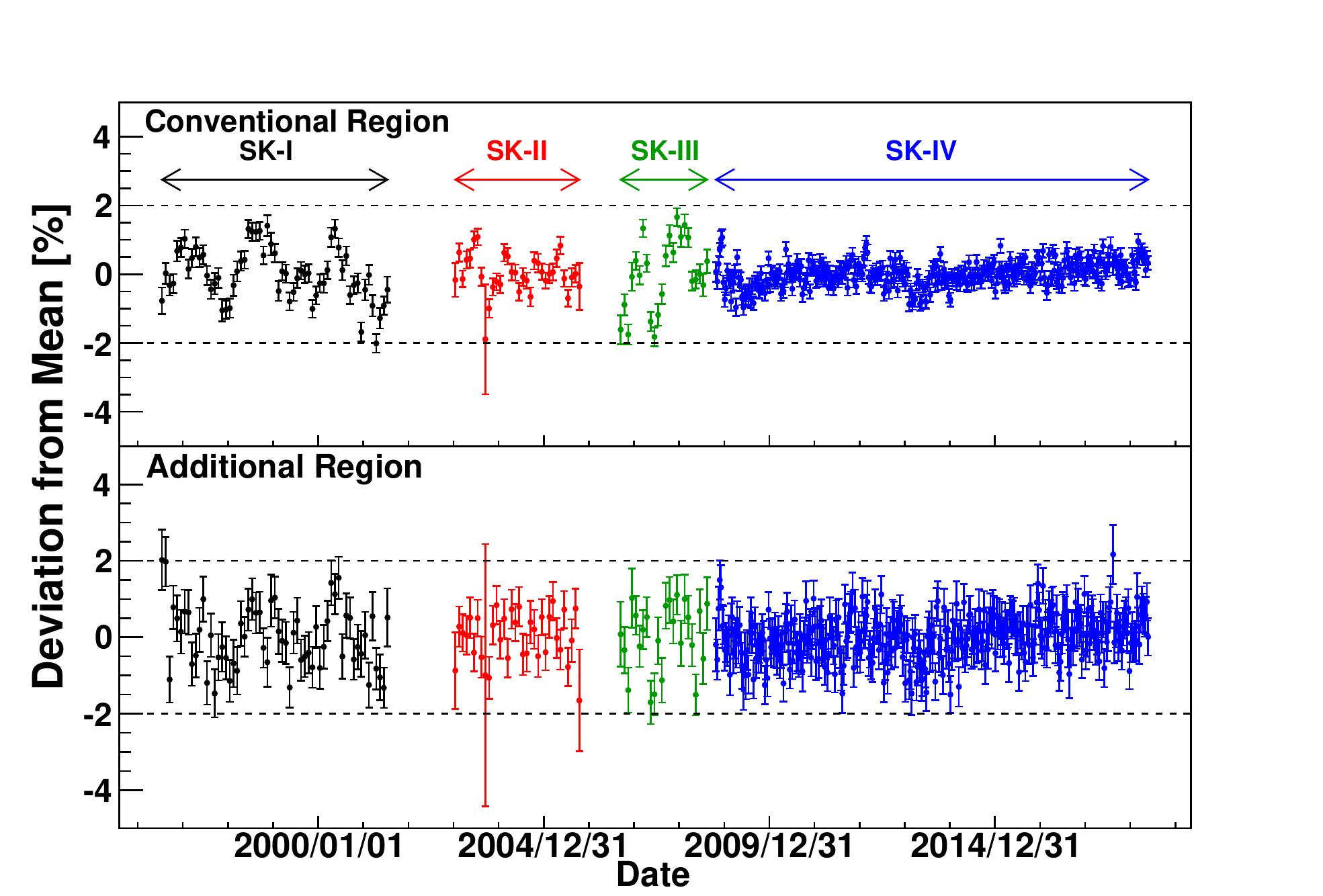}
\caption{Time variation of the average momentum of Michel electrons as a function of date.
For each fiducial volume and SK phase data points are normalized by their mean value. 
The upper (lower) plot corresponds to the conventional (additional) fiducial volume. 
Vertical error bars denote the statistical uncertainty.
For \mbox{SK-I} to \mbox{-III}, each data point corresponds to a one month average and for \mbox{SK-IV}, each data point corresponds to a 10 day average.}
\label{fig:timevari}
\end{figure}

The evolution of the energy scale over time is estimated using the variation in the average 
reconstructed momentum of Michel electrons and the variation in the reconstructed muon momentum over range.
Figure~\ref{fig:timevari} shows the time variation of Michel electron momentum as a function of date.
The time variation is defined as the sample standard deviation of the data value over the run time in each period.
In the end, the total energy scale uncertainty in each SK phase and fiducial volume 
 is the sum in quadrature of this variation with the absolute uncertainty.
A zenith-angle-dependent uncertainty in the energy scale is also estimated using Michel electrons. 
These results are summarized in Table~\ref{tab:escasum} and confirm that 
the data and MC agree to within a few percent in both the conventional and additional fiducial volumes.
These are used as the source of the systematic uncertainty for the proton decay searches described in Sec.~\ref{results}.

\begin{table}[htbp]
\centering
\caption{Summary of the energy scale uncertainty and zenith angle dependent non-uniformity of the energy scale for both the conventional and additional fiducial volumes in units of~\%.
The row ``(Abs., Var.)'' is the breakdown of the energy scale uncertainty, denoting the absolute energy scale uncertainty (Abs.) and the time variation of the energy scale (Var.), respectively.
}
\begin{tabular}{lcccc}  \hline  \hline
\multicolumn{5}{c}{Energy scale uncertainty}  \\
Region &\hspace{0.5mm} SK-I \hspace{0.5mm}&\hspace{0.5mm} SK-II\hspace{0.5mm} &\hspace{0.5mm} SK-III \hspace{0.5mm}&\hspace{0.5mm} SK-IV \hspace{0.5mm} \\
Conventional & 3.3 & 2.0 & 2.4 & 2.1 \\
(Abs., Var.) & (3.1, 0.9) & (2.0, 0.6) & (1.6, 1.8) & (2.1,0.4) \\
\rule[0mm]{0mm}{3mm} \hspace {-2mm} Additional & 3.3 & 3.9 & 2.4 & 2.2 \\
(Abs., Var.) & (3.1, 0.9) & (3.9, 0.6) & (1.6, 1.8) & (2.1,0.6) \\ \hline
\multicolumn{5}{c}{Zenith angle dependent non-uniformity}  \\ 
Conventional & 0.6 & 1.1 & 0.6 & 0.5 \\
Additional & 1.4 & 1.5 & 1.3 & 0.4 \\ \hline \hline
\end{tabular}
\label{tab:escasum}
\end{table}

The fiducial volume for the analysis presented below 
is chosen to be 100~cm $<dwall$ since in this region 
the non-neutrino background contamination is within 1\% 
and the energy scale uncertainty is comparable to the conventional fiducial volume. 
Moving closer to the ID wall would incur larger backgrounds and systematic errors that would degrade the search sensitivity more 
than the search would benefit from the increased exposure.
This enlarged fiducial volume is used in all data in this paper and results in an additional 78~kton$\cdot$year exposure 
analyzed here for the first time. 
The sensitivity improvement with the larger fiducial volume is described in Sec.~\ref{perform}.

\section{Simulation}
\label{simu}
In this analysis $p\rightarrow e^+\pi^0$ and $p\rightarrow \mu^+\pi^0$ MC are used to estimate the signal selection efficiency 
and atmospheric neutrino MC is used to estimate the expected number of background events. 
When generating proton decay events, the decay is assumed to happen with equal probability 
for each proton in the water molecule. 
Hydrogen nuclei (free protons) are stationary and do not interact with other nucleons, 
whereas protons in oxygen nuclei (bound protons) are subject to the effects of the Fermi motion, 
nuclear binding energy, and correlated momentum effects with the surrounding nucleons all of which must be considered 
during their decays.
Furthermore, the interaction of pions with the nuclear medium and the emission of gamma rays and neutrons as the residual 
$^{15}$N nucleus deexcites must also be considered in bound proton decays. 

The Fermi momentum of nucleons in $^{16}$O is simulated based on the electron-${\rm ^{12}C}$ scattering experiment data~\cite{fermi}.
For such decays the effect of the nuclear binding energy is introduced 
as an effective mass of the proton, $M^{\prime}_P = M_P - E_b$, where $M^{\prime}_P$ is the modified proton mass, $M_P$ is the proton rest mass and $E_b$ is the nuclear binding energy.
Here $E_b$ is drawn from a Gaussian distribution with a mean of 39.0~MeV (15.5~MeV) and a standard deviation of 10.2~MeV(3.8~MeV) for $S$ ($P$) state protons.
The ratio of protons in the $S$ state to those in the $P$ state is taken to be 1:3 based on the nuclear shell model~\cite{shell}.
Proton decay kinematics can be distorted by repeated collisions with surrounding nuclei during their decay, an effect known as 
correlated decay, with a predicted probability of about 10\%~\cite{correlated}.
Neutral pion absorption, scattering, and charge exchange processes (final state interactions, FSI) are 
simulated with the NEUT cascade model~\cite{neut,fsi}.
Detailed descriptions about $\pi$-FSI and its interaction breakdown plot can be found in the last paper~\cite{last}.
Gamma ray and neutron emission following a bound proton decay is simulated based on~\cite{ejiri}, 
where the latter has a probability of less than 10\%.

\begin{figure*}[htbp]
\centering
\includegraphics[width=1.0\textwidth, clip, viewport =  0.00   0.00   567.00   192.00]{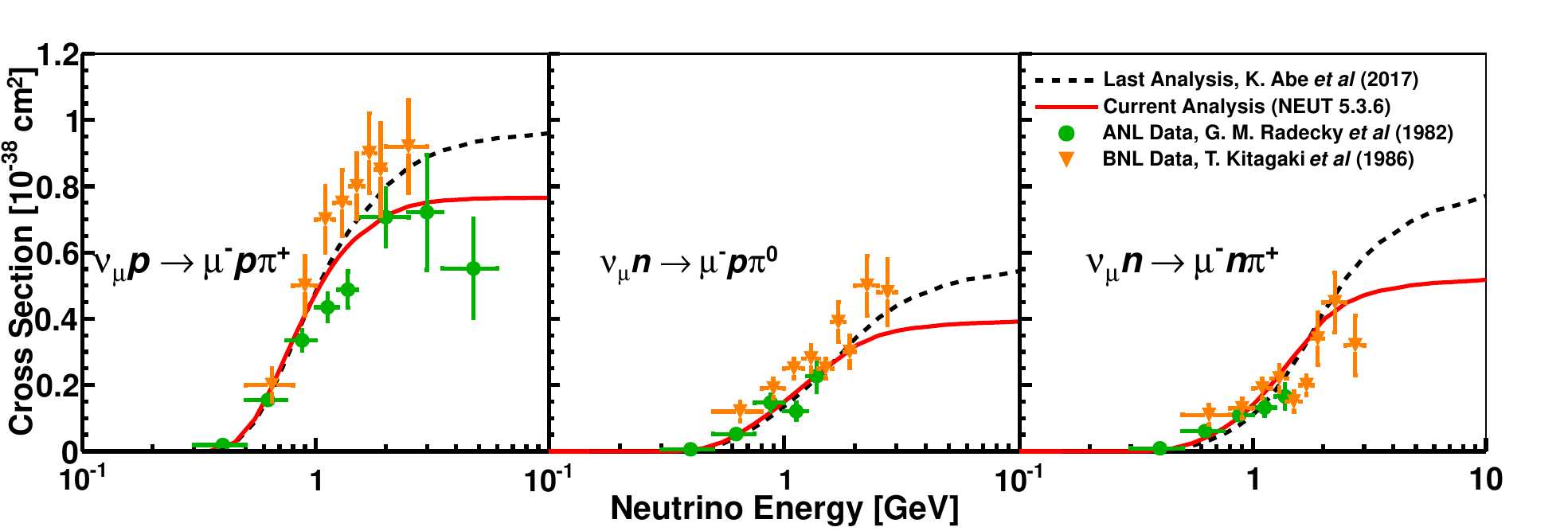}
\caption{Charged current single $\pi$ production cross section as a function of neutrino energy. The dotted black line corresponds to the model used in previous SK proton decay searches, NEUT 5.1.4, and the solid red line shows that for the current analysis, NEUT 5.3.6.}
\label{fig:cc1pi}
\end{figure*}

Atmospheric neutrinos are simulated using the HKKM flux~\cite{honda} and NEUT~\cite{neut}.
The neutrino interaction model has been updated since the last paper~\cite{last} and 
a summary of each interaction mode update can be found in~\cite{osc}.
For this analysis the update to the charged current single $\pi$ production model 
is the most important as it is the dominant background interaction mode.
Previously this interaction was simulated using the Rein-Sehgal model~\cite{reinsehgal} 
but updated form factors have been obtained from a simultaneous fit~\cite{graczyk} to neutrino scattering data from bubble chamber 
experiments~\cite{bubble} and are included in the new model.
A comparison of the cross section as a function of neutrino energy between the previous model (NEUT 5.1.4) 
and the current model (NEUT 5.3.6) is shown in Figure~\ref{fig:cc1pi}.
Since neutrinos of about 1 to 3~GeV are the dominant background to proton decay, this change is expected to 
reduce the number of background events by about 15\%.
At the same time the neutrino-nucleon momentum transfer is reduced in the new model, which subsequently increases
 the momentum of the produced lepton. 
As the search below focuses on a back-to-back event topology with a low total momentum characteristic of proton decays,
this change in the momentum transfer is expected to further reduce backgrounds by 15\%.
In total a 30\% reduction of the background is expected with the new model.

Neutrons play an important role in distinguishing proton decay events from atmospheric neutrino backgrounds. 
The dominant production process for the latter, representing 70\% of produced neutrons, 
is via the interaction of secondary hadrons with water.
These processes are simulated with the CALOR package~\cite{calor}, which uses HETC~\cite{hetc} for hadrons below 10~GeV, 
FLUKA~\cite{fluka} for hadrons above 10~GeV, and MICAP~\cite{micap} for neutrons below 20~MeV.
Low energy background sources, such as gamma rays from the radioactive decay of nuclei in the detector material, are not simulated in the MC 
but have non-negligible impacts on the search for 2.2~MeV gamma rays from neutron capture on hydrogen.
Therefore, PMT information from random trigger data is added to the MC 18~$\mu$sec after the primary event trigger 
to model these backgrounds. 

\section{Search Method}
\label{perform}

The event selection criteria for the search for $p\rightarrow e^+\pi^0$ and $p\rightarrow \mu^+\pi^0$ modes are as follows:
\begin{description}
\item[C1] Events must pass the updated FC event selection criteria (see Sec.~\ref{enlarge}) with a vertex within the fiducial volume.
\item[C2] Events must have two or three reconstructed Cherenkov rings.
\item[C3] All rings must be reconstructed as showering for $p\rightarrow e^+\pi^0$ and exactly one ring must be a non-showering for $p\rightarrow \mu^+\pi^0$.
\item[C4] There must be no tagged Michel electrons for $p\rightarrow e^+\pi^0$ and exactly one for $p\rightarrow \mu^+\pi^0$.
\item[C5] For events with three rings, the reconstructed $\pi^0$ mass must be $85 < M_{\pi^0} < 185 ~{\rm MeV}/c^2$.
\item[C6] The total reconstructed mass must be $800 < M_{{\rm tot}} < 1050 ~{\rm MeV}/c^2$.
\item[C7] The total momentum must be $P_{{\rm tot}} < 250 ~{\rm MeV}/c$.
\item[C8] For the \mbox{SK-IV} data there must be no tagged neutrons.
\end{description} \par

\begin{figure*}[htbp]
\begin{center}
\begin{tabular}{c}
\begin{minipage}{0.50\hsize}
\begin{center}
\includegraphics[width=1.0\textwidth, clip, viewport = 0.00   0.00   567.00   550.00]{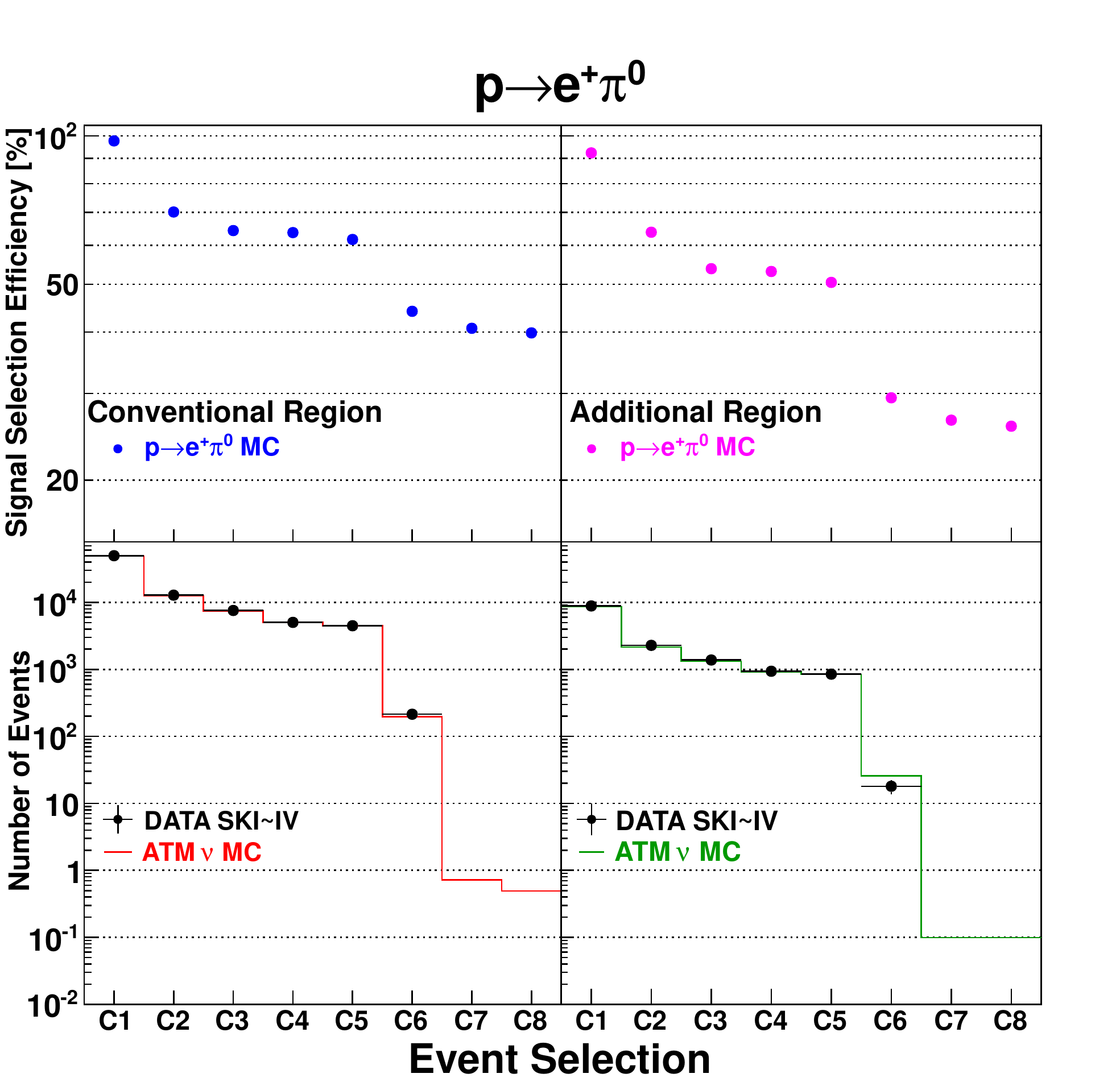}
\end{center}
\end{minipage}
\begin{minipage}{0.50\hsize}
\begin{center}
\includegraphics[width=1.0\textwidth, clip, viewport = 0.00   0.00   567.00   550.00]{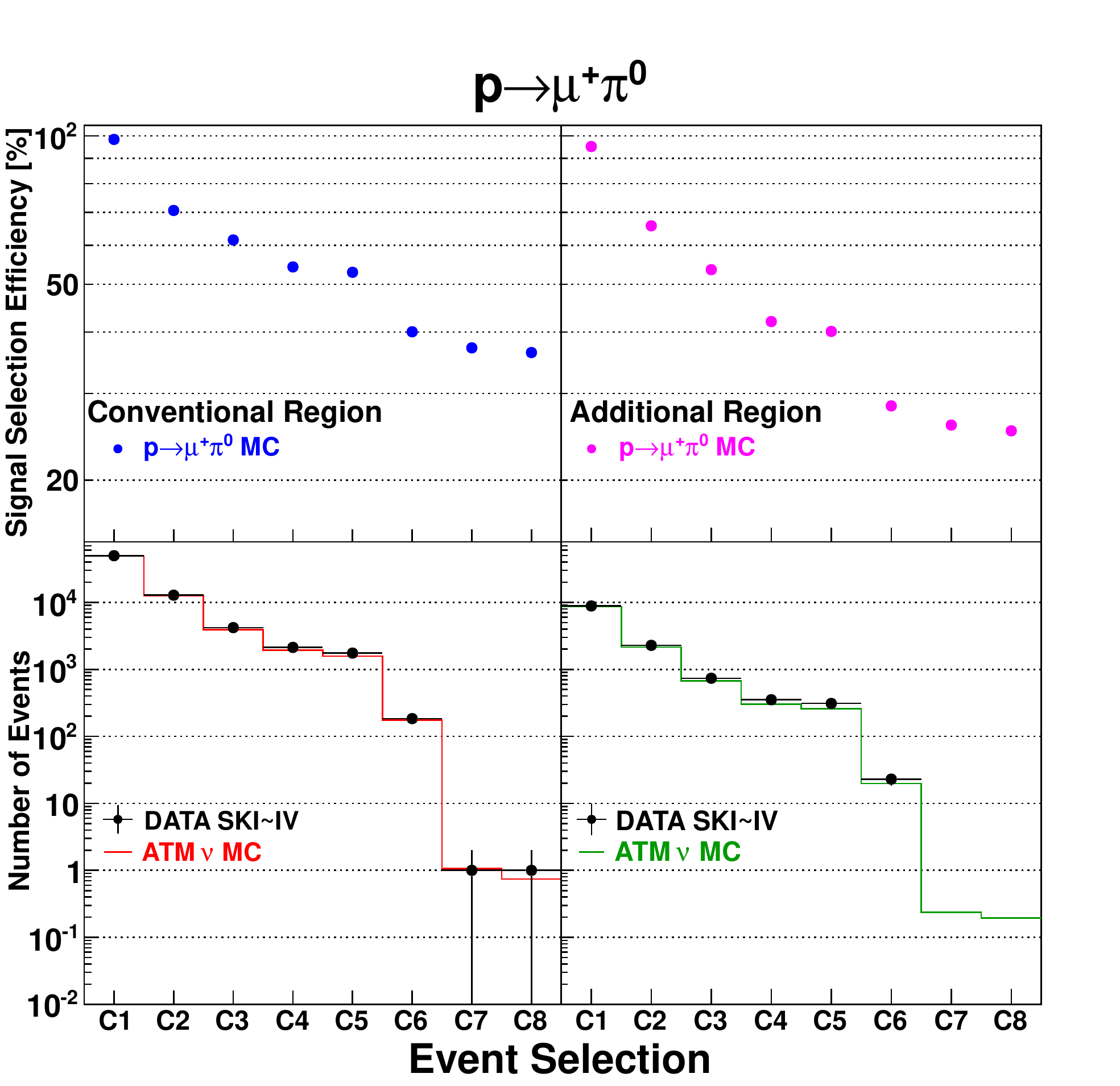}
\end{center}
\end{minipage}
\end{tabular}
\caption{The signal selection efficiencies (upper) and the expected number of atmospheric neutrino background events (lower, histogram) and data candidates (lower, black dots) for $p\rightarrow e^+\pi^0$ (left) and $p\rightarrow \mu^+\pi^0$ (right). 
Vertical error bars on the data points denote the statistical uncertainty. 
In each plot, the left panel corresponds to the conventional fiducial volume and the right panel to the additional fiducial volume. 
Atmospheric neutrino MC is normalized by livetime and includes reweighting to the latest SK oscillation fit~\cite{osc}. 
The combined data from \mbox{SK-I} to \mbox{-IV} is shown along with the combined signal and background MC.
} 
\label{fig:cut}
\end{center}
\end{figure*}

For criterion {\bf C5} in the $p\rightarrow e^+\pi^0$ mode, the $\pi^{0}$ mass is calculated with every pair of rings.
The pair giving the $\pi^{0}$ mass closest to 135~${\rm MeV}/c^2$ is considered to be two gamma rays from the $\pi^0$ decay.
The signal selection efficiencies and the expected number of atmospheric neutrino background events for both the conventional and additional fiducial volumes are estimated with MC and are shown at each step of the selection in Figure~\ref{fig:cut}.
The expected number of atmospheric neutrino background events is estimated using a 500-year equivalent exposure of atmospheric neutrino MC events for each SK phase (2000~years in total) and is normalized to detector livetime and the latest SK oscillation result~\cite{osc}.
Signal selection efficiencies in the additional fiducial volume have been improved by about 20\% based on improvements 
in the event reconstruction algorithm described in Section~\ref{enlarge}.
However, due to fewer hit PMTs and a higher likelihood of particles escaping the ID,
the efficiencies are still lower than those in the conventional fiducial volume.
This is especially true for {\bf C3}, where the PID is degraded due to reduced hits, and {\bf C6}, where energy is lost due to an escaping particle, for $p\rightarrow e^+\pi^0$.
The situation is similar for $p\rightarrow \mu^+\pi^0$, where losses are seen at {\bf C3} (PID) and also at {\bf C4} due to the muon exiting the ID.

Two signal regions are defined: a lower total momentum region ($P_{\rm tot} < 100 ~{\rm MeV}/c$) and a higher total momentum region ($100 \leq P_{\rm tot} < 250 ~{\rm MeV}/c$).
The final signal selection efficiencies and expected number of atmospheric neutrino background events are summarized in Table~\ref{tab:summary}.
For the $p\rightarrow e^+\pi^0$ ($p\rightarrow \mu^+\pi^0$) mode, the livetime-weighted total signal selection efficiency 
and expected number of background events in the additional fiducial volume are 25.8\% (25.2\%) and 0.10 (0.19), respectively, while in the conventional fiducial volume they are 39.8\% (36.3\%) and 0.49 (0.74).
The expected background rate for $p\rightarrow e^+\pi^0$ without {\bf C8} is 1.83~/Mt$\cdot$years and 
is consistent with previous estimations by the K2K 1~kton water Cherenkov detector of 1.63$^{+0.42}_{-0.33}$(stat)$^{+0.45}_{-0.51}$(sys)~/Mt$\cdot$years~\cite{k2k}.
A breakdown of the remaining background events by interaction mode is shown in Table~\ref{tab:nuint}. 
There are no significant differences in the dominant charged current single $\pi$ production backgrounds in the two fiducial volumes.
Although the signal selection efficiencies in the additional fiducial volume are lower than in the conventional fiducial volume, 
the enlarged fiducial volume leads to an increase in the search sensitivity at the 90\% C.L. of about 12\%. 

\begin{table*}[htbp]
\centering
\caption{Summary of signal selection efficiencies, the expected number of background events and the number of data candidates for 92.1 (19.3), 49.1 (10.3), 31.9 (6.7) and 199.5 (41.8)~kton$\cdot$years exposures from the conventional (additional) fiducial volume of \mbox{SK-I}, \mbox{-II}, \mbox{-III}, and \mbox{-IV}.
The ``Enlarged'' row shows results for the enlarged 27.2~kton fiducial volume.
Here ``Lower'' and ``Upper'' indicate $P_{\rm{tot}} < 100 ~{\rm MeV}/c$ and $100 \leq P_{\rm{tot}} < 250 ~{\rm MeV}/c$, respectively. 
Errors in the signal selection efficiency and the expected number of background events are the quadratic sum of 
MC statistical error and systematic errors.}
\begin{tabular}{lccccccccccccccc} \hline  \hline
Search Mode & & \multicolumn{4}{c}{Signal Selection Efficiency (\%)} & \hspace{0mm} & \multicolumn{4}{c}{Background (events)} & \hspace{0mm} & \multicolumn{4}{c}{Candidate (events)}  \\
Region & & I & II & III & IV & \hspace{0mm} & I & II & III & IV & \hspace{0mm}  & I & II & III & IV \\ \hline
$p\rightarrow e^+\pi^0$  \\
\multirow{2}{*}{Conventional} & (Lower) & 19.9$\pm$1.9 & 18.1$\pm$1.8 & 20.3$\pm$1.8 & 19.6$\pm$1.6 & \hspace{0mm} & $<0.01$ & 0.01$\pm$0.01 & $<0.01$ & $<0.01$ & \hspace{0mm}  & 0 & 0 & 0 & 0 \\
 & (Upper) & 21.0$\pm$3.5 & 20.2$\pm$3.2 & 21.1$\pm$3.5 & 19.8$\pm$3.3 & \hspace{0mm} & 0.13$\pm$0.05 & 0.11$\pm$0.04 & 0.05$\pm$0.02 & 0.20$\pm$0.09 & \hspace{0mm}  & 0 & 0 & 0 & 0 \\
\multirow{2}{*}{Additional} & (Lower) & 9.6$\pm$1.5 & 8.8$\pm$1.4 & 9.9$\pm$1.7 & 11.0$\pm$1.5 & \hspace{0mm} & 0.01$\pm$0.01 & $<0.01$ & $<0.01$ & $<0.01$ & \hspace{0mm}  & 0 & 0 & 0 & 0 \\
 & (Upper) & 14.5$\pm$2.9 & 14.9$\pm$2.7 & 16.4$\pm$2.8 & 15.9$\pm$2.6 & \hspace{0mm} & 0.02$\pm$0.01 & $<0.01$ & 0.02$\pm$0.01 & 0.05$\pm$0.04 & \hspace{0mm}  & 0 & 0 & 0 & 0 \\
\rule[0mm]{0mm}{6mm} \hspace {-2mm} \multirow{2}{*}{Enlarged} & (Lower) & 18.3$\pm$1.7 & 16.6$\pm$1.7 & 18.7$\pm$1.7 & 18.2$\pm$1.5 & \hspace{0mm} & 0.01$\pm$0.01 & 0.01$\pm$0.01 & $<0.01$ & $<0.01$ & \hspace{0mm}  & 0 & 0 & 0 & 0 \\
 & (Upper) & 20.0$\pm$3.3 & 19.4$\pm$3.0 & 20.3$\pm$3.3 & 19.2$\pm$3.1 & \hspace{0mm} & 0.15$\pm$0.06 & 0.11$\pm$0.04 & 0.07$\pm$0.03 & 0.25$\pm$0.11 & \hspace{0mm}  & 0 & 0 & 0 & 0 \\ \hline
$p\rightarrow \mu^+\pi^0$   \\
\multirow{2}{*}{Conventional} & (Lower) & 17.0$\pm$1.6 & 16.2$\pm$1.5 & 17.5$\pm$1.6 & 19.9$\pm$1.9 & \hspace{0mm} & 0.03$\pm$0.02 & $<0.01$ & 0.01$\pm$0.01 & $<0.01$ & \hspace{0mm}  & 0 & 0 & 0 & 0 \\
 & (Upper) & 16.7$\pm$3.1 & 16.5$\pm$2.8 & 16.8$\pm$3.0 & 18.9$\pm$3.7 & \hspace{0mm} & 0.19$\pm$0.06 & 0.10$\pm$0.04 & 0.06$\pm$0.02 & 0.34$\pm$0.12 & \hspace{0mm}  & 0 & 0 & 0 & 1 \\
\multirow{2}{*}{Additional} & (Lower) & 11.1$\pm$1.5 & 8.8$\pm$1.2 & 11.0$\pm$1.4 & 12.7$\pm$1.2 & \hspace{0mm} & $<0.01$ & $<0.01$ & $<0.01$ & $<0.01$ & \hspace{0mm}  & 0 & 0 & 0 & 0 \\
 & (Upper) & 12.0$\pm$2.3 & 12.6$\pm$2.3 & 12.5$\pm$2.2 & 14.7$\pm$2.6 & \hspace{0mm} & 0.02$\pm$0.02 & 0.03$\pm$0.01 & 0.02$\pm$0.01 & 0.12$\pm$0.06 & \hspace{0mm}  & 0 & 0 & 0 & 0 \\ 
\rule[0mm]{0mm}{6mm} \hspace {-2mm} \multirow{2}{*}{Enlarged} & (Lower) & 16.0$\pm$1.5 & 14.9$\pm$1.4 & 16.4$\pm$1.5 & 18.7$\pm$1.7 & \hspace{0mm} & 0.03$\pm$0.02 & 0.01$\pm$0.01 & 0.01$\pm$0.01 & $<0.01$ & \hspace{0mm}  & 0 & 0 & 0 & 0 \\
 & (Upper) & 16.0$\pm$2.9 & 15.8$\pm$2.7 & 16.1$\pm$2.9 & 18.2$\pm$3.4 & \hspace{0mm} & 0.21$\pm$0.07 & 0.14$\pm$0.04 & 0.08$\pm$0.03 & 0.46$\pm$0.15 & \hspace{0mm}  & 0 & 0 & 0 & 1 \\ \hline \hline
\end{tabular}  
\label{tab:summary}
\end{table*}

\begin{table}[htbp]
\centering
\caption{Breakdown of interaction modes for background events remaining in the signal region for the $p\rightarrow e^+\pi^0$ and $p\rightarrow \mu^+\pi^0$ searches in units of~\%. 
Here, CC and NC stand for charged-current and neutral-current, respectively and QE, 1$\pi$ and DIS stand for quasi-elastic scattering, single $\pi$ production and deep inelastic scattering, respectively.}
\begin{tabular}{lccccc}  \hline  \hline
$p\rightarrow e^+\pi^0$  \\
Region &\hspace{0.5mm} CCQE \hspace{0.5mm}&\hspace{0.5mm} CC1$\pi$\hspace{0.5mm} &\hspace{0.5mm} CCDIS \hspace{0.5mm}&\hspace{0.5mm} NC1$\pi$ \hspace{0.5mm}&\hspace{0.5mm} NCDIS\hspace{0.5mm} \\
Conventional & 18 & 63 & 10 & 1 & 8 \\
Additional & 21 & 62 & 14 & 0 & 3 \\ \hline
$p\rightarrow \mu^+\pi^0$  \\ 
Conventional & 13 & 65 & 18 & 0 & 4 \\
Additional & 7 & 60 & 17 & 0 & 16 \\ \hline \hline
\end{tabular}
\label{tab:nuint}
\end{table}

\begin{figure*}[htbp]
\centering
\includegraphics[width=1.0\textwidth, clip, viewport = 0.000000 0.000000 2363 1130]{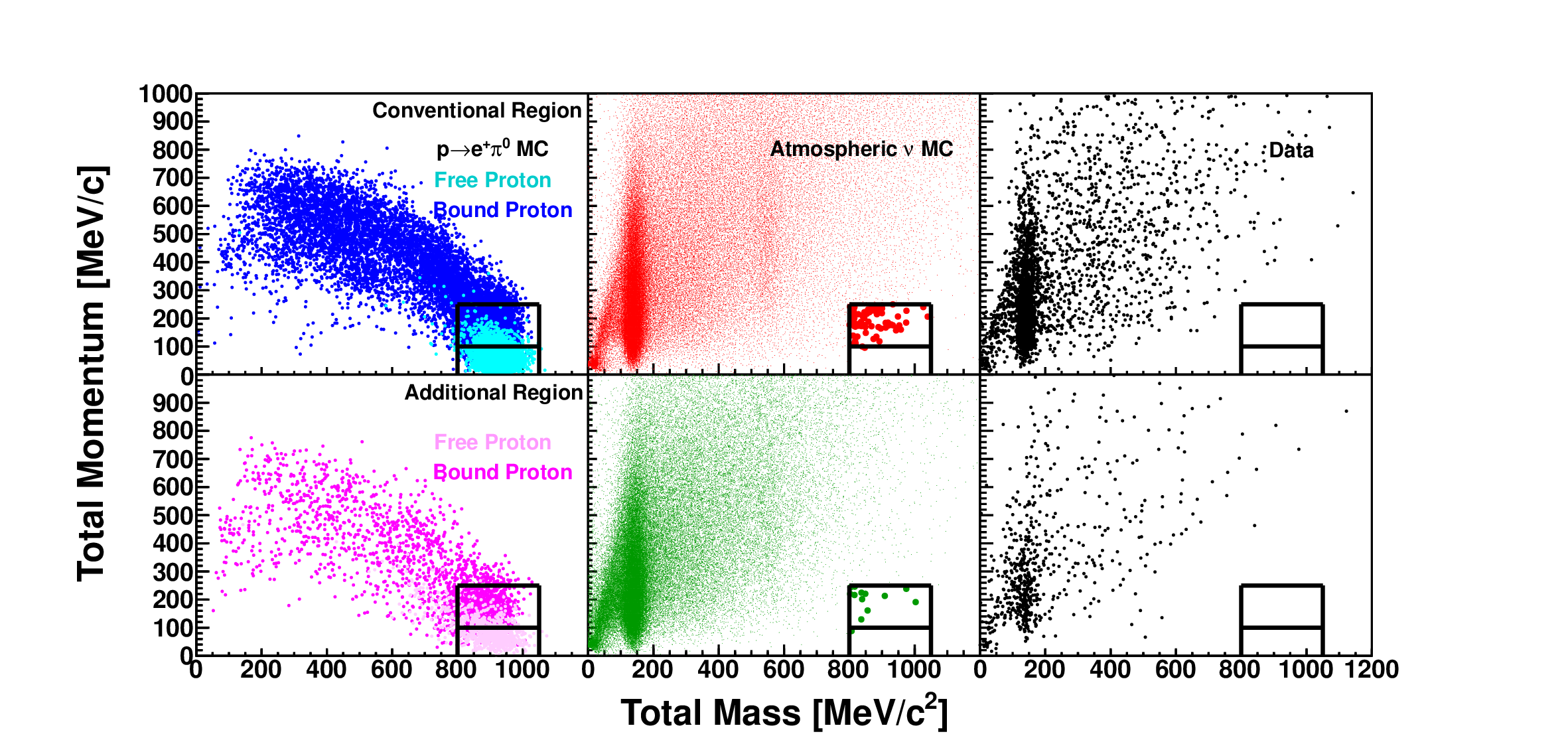}
\caption{Reconstructed total mass shown against the total momentum distributions for $p\rightarrow e^+\pi^0$ after all 
cuts except those on these variables. 
The top plots correspond to the conventional and the bottom plots correspond to the additional fiducial volume.
The left panels show the signal MC (\mbox{SK-I} to \mbox{-IV} are combined), where lighter colors show 
free protons and dark colors show bound protons. 
The middle panels show the 2000~year-equivalent atmospheric neutrino MC.
The right panels show all the combined data \mbox{SK-I} to \mbox{-IV}. 
The black box shows the signal region and for the middle panels the markers in the signal region have been enlarged for visibility.}
\label{fig:2depi0}
\end{figure*}

\begin{figure*}[htbp]
\centering
\includegraphics[width=1.0\textwidth, clip, viewport = 0.000000 0.000000 2363 1130]{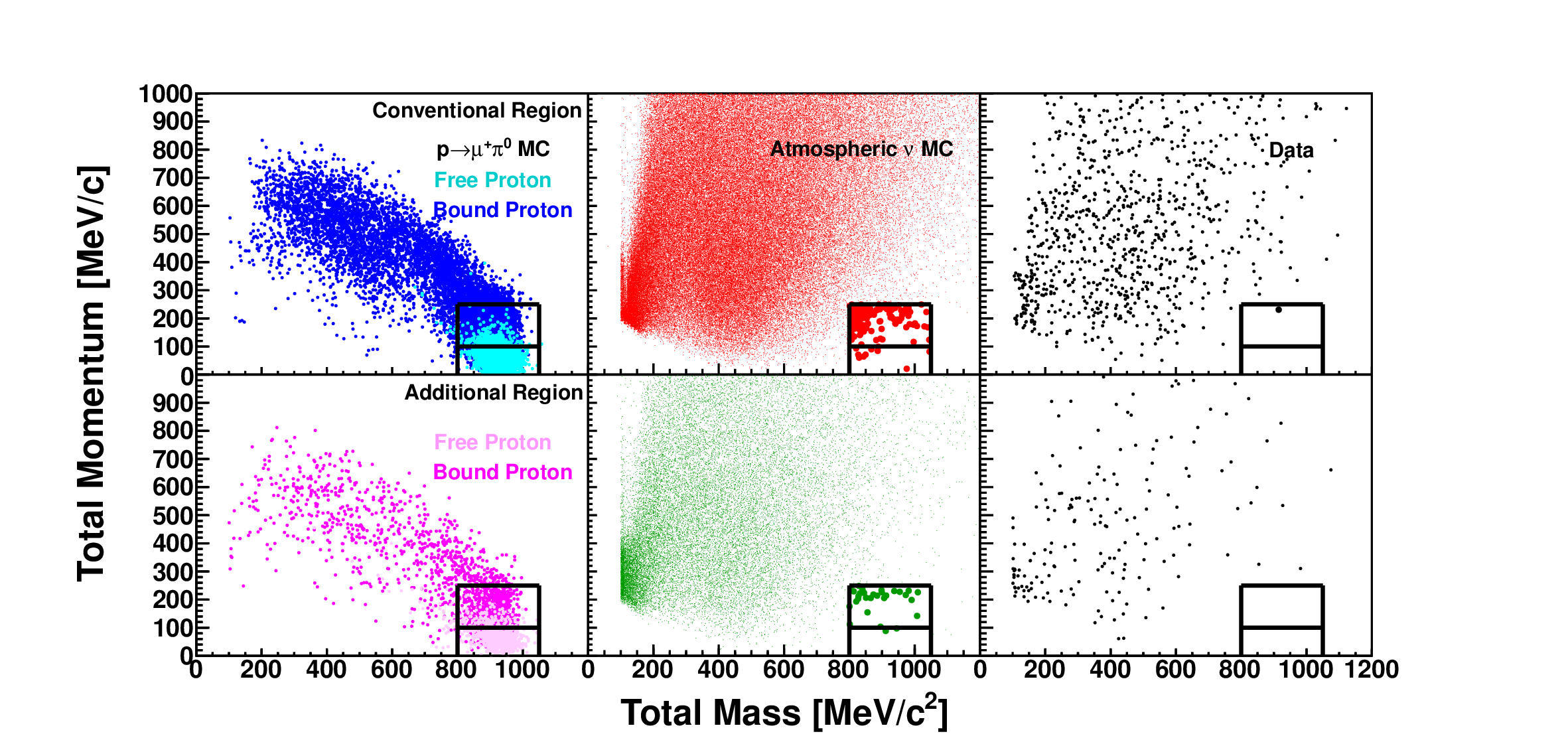}
\caption{Reconstructed total mass shown against the total momentum distributions for $p\rightarrow \mu^{+}\pi^0$ after all 
cuts except those on these variables. 
The top plots correspond to the conventional and the bottom plots correspond to the additional fiducial volume.
The left panels show the signal MC (\mbox{SK-I} to \mbox{-IV} are combined), where lighter colors show 
free protons and dark colors show bound protons. 
The middle panels show the 2000~year-equivalent atmospheric neutrino MC.
The right panels show all the combined data \mbox{SK-I} to \mbox{-IV}. 
The black box shows the signal region and for the middle and right panels the markers in the signal region have been enlarged for visibility.}
\label{fig:2dmupi0}
\end{figure*}

\begin{figure*}[htbp]
\begin{center}
\begin{tabular}{c}
\begin{minipage}{0.50\hsize}
\begin{center}
\includegraphics[width=1.0\textwidth, clip, viewport = 0.00   0.00   567.00   416.00]{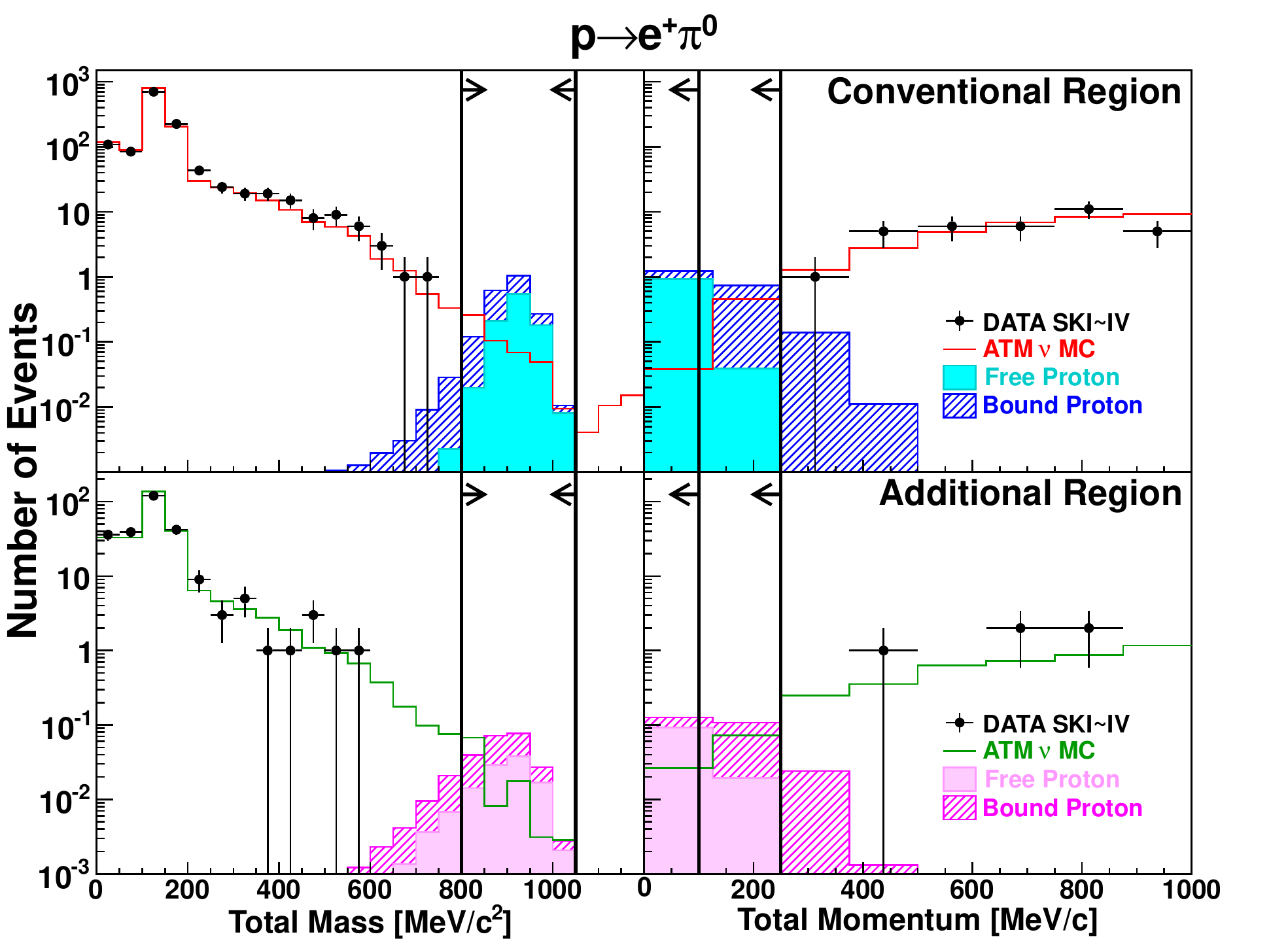}
\end{center}
\end{minipage}
\begin{minipage}{0.50\hsize}
\begin{center}
\includegraphics[width=1.0\textwidth, clip, viewport = 0.00   0.00   567.00   416.00]{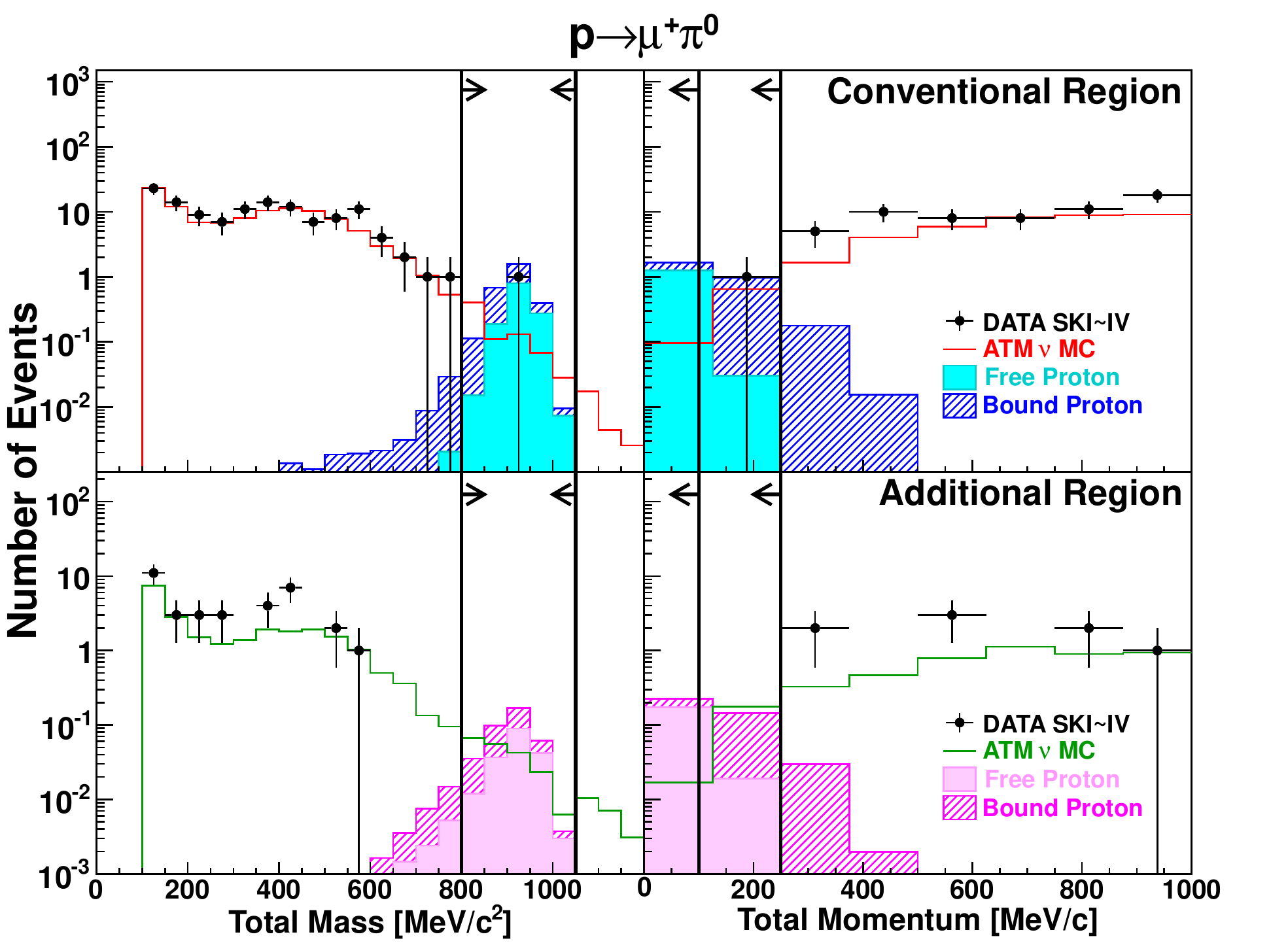}
\end{center}
\end{minipage}
\end{tabular}
\caption{Reconstructed total mass and total momentum distributions for $p\rightarrow e^+\pi^0$ (left) and $p\rightarrow \mu^+\pi^0$ (right)
 after all the cuts except those on the plotted variables. 
The top panels show the conventional and the bottom panels show the additional fiducial volume.
The left panels show the reconstructed total mass and the right panels show the reconstructed total momentum distributions. 
The signal MC histograms are stacked, showing free proton decay events (light color) and bound proton decay events (dark color, hatched). 
The combined result from \mbox{SK-I} to \mbox{-IV} is shown normalized by the 90\% C.L. upper limit on the signal derived in this work. 
Atmospheric neutrino MC (red and green) is normalized by livetime and includes reweighting to the latest SK oscillation fit~\cite{osc}.
Vertical error bars on the data points denote the statistical uncertainty. Bold lines and arrows show the signal region.}
\label{fig:1dplot}
\end{center}
\end{figure*}

Figures~\ref{fig:2depi0} and~\ref{fig:2dmupi0} show the two-dimensional total mass and total momentum distributions 
 for the signal MC, the atmospheric neutrino MC, and all data from \mbox{SK-I} to \mbox{-IV} after all the selection cuts have been applied 
except the cuts on the plotted variables.
The lower $P_{{\rm tot}}$ signal region contains mostly decays from free protons and is nearly background-free in both fiducial volumes.
One-dimensional distributions of each variable are shown in Figure~\ref{fig:1dplot} after all the selection cuts except the cut on the plotted variable.
As discussed in Sec.~\ref{enlarge}, the total mass and total momentum distributions in the additional fiducial volume are wider than those in the conventional due to the effects of lower numbers of hit PMTs and particles escaping the ID.

\section{Search Results}
\label{results}
Applying the search criteria above to a 450~kton$\cdot$years exposure of the data 
resulted in no signal candidates for the $p\rightarrow e^+\pi^0$ mode in either the conventional or additional fiducial volume.
A single candidate was found in the $p\rightarrow \mu^+\pi^0$ search's upper $P_{\rm tot}$ part of the conventional fiducial volume and is the same candidate found 
in the previous search~\cite{last}.
No additional candidates were found in the additional fiducial volume.
The number of data candidates is not significantly higher than the expected number of atmospheric neutrino events 
in either mode,  for $p\rightarrow e^+\pi^0$ 0.59 events and for $p\rightarrow \mu^+\pi^0$ 0.94 events in total. 
For the latter the Poisson probability to observe 1 event or more with mean value of 0.94 is 60.9\%.
The data are consistent with the atmospheric neutrino MC prediction in and outside of the signal regions. 

Systematic uncertainties for the selection efficiencies and the expected number of background events are summarized in Tables~\ref{tab:sigsys} 
and~\ref{tab:bgsys}.
Uncertainties in the vertex position, the number of identified Cherenkov rings, the identified particle type, the number of tagged Michel electrons, the energy scale, and the number of tagged neutrons are considered as uncertainties in the reconstruction and are evaluated in each fiducial volume separately.
For the signal selection efficiency, the uncertainties associated with the reconstruction in the additional fiducial volume are relatively higher than those in the conventional fiducial volume.
As mentioned above, the reconstruction uncertainty is dominated by uncertainty in the energy scale.
Due to the degraded momentum resolution and the effects of escaping particles, more signal MC events in the additional fiducial volume 
are reconstructed near the total mass and total momentum cut boundaries than in the conventional one (see Figures~\ref{fig:epi0tmass3} and~\ref{fig:1dplot}).
This has the effect of making the selection efficiency in the additional fiducial volume more sensitive to uncertainties in the energy scale.

Concerning the modeling of physics processes, uncertainties in the Fermi momentum, the 
$\pi$ FSI, and correlated decays are considered.
The same source uncertainties are used in both fiducial volumes. 
Because of the worse momentum resolution in the additional fiducial volume, more free proton decay events are reconstructed in the higher $P_{\rm{tot}}$ signal region and the fraction of the bound proton decay events is lower than in the conventional fiducial volume.
Since all considered uncertainties have impacts only on the bound proton decay events, this smaller bound proton decay event fraction leads smaller uncertainties in the additional fiducial volume and higher $P_{\rm{tot}}$ signal region for every physics model item.
Furthermore, it should be noted that the uncertainty in the Fermi momentum distribution manifests as changes in the total reconstructed momentum 
and can be aggravated by the reconstruction performance in each fiducial volume. 
This uncertainty is estimated based on the initial $^{16}$O proton momentum distribution in models based on data (see Sec.~\ref{simu}) in comparison 
with the NEUT's Fermi gas model.
The degraded momentum reconstruction performance in the additional fiducial volume smears the effect from this model change and 
results in relatively small uncertainties in the higher $P_{{\rm tot}}$ signal region.

Uncertainties affecting the expected number of atmospheric neutrino background events 
are evaluated combining the upper and lower $P_{\rm tot}$ signal regions because there are only a few events 
remaining in the latter. 
The uncertainties associated with the event reconstruction are evaluated using the same methods as the signal selection efficiency.
The systematic uncertainty associated with the vertex position, which is sub-dominant in the reconstruction items and estimated by artificially shifting the vertex position, partially anticorrelates between the conventional and additional fiducial volume, resulting in the smallest uncertainty after combining the two fiducial volumes.
Physics model uncertainties include those from the neutrino flux, the neutrino interaction (cross section) model, the $\pi$ FSI model, and $\pi$ secondary interactions (SI) in water.
Other than the reconstruction uncertainties discussed above, those from the neutrino interaction model are dominant.
As shown in Table~\ref{tab:nuint} and Figure~\ref{fig:1dplot}
the background distributions and uncertainties are similar in the two fiducial volumes.
The systematic uncertainty associated with the charged current single $\pi$ production 
is assigned as the change in the background after varying the form factors, introduced in Sec.~\ref{simu}, by their error size.
In the additional (conventional) fiducial volume this results in a 
9.1\% (11.8\%) change in the  $p\rightarrow e^+\pi^0$ background and a
9.1\% (13.9\%) change for  $p\rightarrow \mu^+\pi^0$.
In the previous analysis this uncertainty was 7.9\% and 5.3\% for the conventional fiducial volume of the two modes, respectively. 
In addition to the above, an uncertainty on the detector exposure is conservatively estimated to be 1\%.

\begin{table*}[htbp]
\centering
\caption{Summary of systematic uncertainties [\%] on the signal selection efficiency for each fiducial volume.
The ``Enlarged'' row shows the result for the combination of the two fiducial volumes. 
Here ``Lower'' and ``Upper'' show the $P_{\rm{tot}} < 100 ~{\rm MeV}/c$ and $100 \leq P_{\rm{tot}} < 250 ~{\rm MeV}/c$ signal regions, respectively.}
\begin{tabular}{lcccccc}  \hline  \hline
Region & & \hspace{2mm} Correlated Decay \hspace{2mm} & \hspace{2mm} Fermi momentum \hspace{2mm} & \hspace{2mm} $\pi$ FSI \hspace{2mm} & \hspace{2mm} Reconstruction \hspace{2mm} & \hspace{2mm} Total \\ \hline
$p\rightarrow e^+\pi^0$  \\
\multirow{2}{*}{Conventional} & (Lower) & 1.7 & 7.2 & 2.8 & 3.5 & 8.6 \\
 & (Upper) & 9.0 & 6.7 & 11.9 & 2.8 & 16.6 \\
\multirow{2}{*}{Additional} & (Lower) & 2.7 & 7.8 & 3.8 & 10.3 & 13.7 \\
 & (Upper) & 8.4 & 1.0 & 10.6 & 10.0 & 16.8 \\
\rule[0mm]{0mm}{6mm} \hspace {-2mm} \multirow{2}{*}{Enlarged} & (Lower) & 1.9 & 7.2 & 2.9 & 3.6 & 8.8 \\
 & (Upper) & 8.9 & 5.7 & 11.7 & 3.2 & 16.1 \\ \hline
$p\rightarrow \mu^+\pi^0$  \\ 
\multirow{2}{*}{Conventional} & (Lower) & 1.9 & 7.3 & 2.7 & 4.5 & 9.2 \\
 & (Upper) & 9.3 & 8.3 & 13.3 & 4.0 & 18.7 \\
\multirow{2}{*}{Additional} & (Lower) & 2.0 & 6.8 & 2.9 & 7.1 & 10.4 \\
 & (Upper) & 8.4 & 4.0 & 11.3 & 9.5 & 17.5 \\
\rule[0mm]{0mm}{6mm} \hspace {-2mm} \multirow{2}{*}{Enlarged} & (Lower) & 1.9 & 7.2 & 2.7 & 4.8 & 9.3 \\
 & (Upper) & 9.2 & 7.7 & 13.0 & 4.5 & 18.2 \\ \hline \hline
\end{tabular}
\label{tab:sigsys}
\end{table*}

\begin{table*}[htbp]
\centering
\caption{Summary of systematic uncertainties [\%] on the expected number of atmospheric neutrino background events for each fiducial volume. 
They have been evaluated for the combined $P_{\rm{tot}} < 100 ~{\rm MeV}/c$ and $100 \leq P_{\rm{tot}} < 250 ~{\rm MeV}/c$ signal regions. 
Here the ``Enlarged'' row shows the result for the combination of the two fiducial volumes.}
\begin{tabular}{lccccc}  \hline  \hline
Region & \hspace{2mm} Neutrino flux \hspace{2mm} & \hspace{2mm} Neutrino interaction \hspace{2mm} & \hspace{2mm} $\pi$ FSI and SI \hspace{2mm} & \hspace{2mm} Reconstruction \hspace{2mm} & \hspace{2mm} Total \hspace{2mm}\\ \hline
$p\rightarrow e^+\pi^0$  \\
Conventional & 7.3 & 21.2 & 12.8 & 21.3 & 33.5 \\
Additional & 7.2 & 17.3 & 13.9 & 25.2 & 34.3 \\
\rule[0mm]{0mm}{6mm} \hspace {-2mm} Enlarged & 7.3 & 19.8 & 12.7 & 20.5 & 32.0 \\ \hline
$p\rightarrow \mu^+\pi^0$  \\ 
Conventional & 7.2 & 19.0 & 9.3 & 16.5 & 27.8 \\
Additional & 7.3 & 16.8 & 11.3 & 21.4 & 30.3 \\
\rule[0mm]{0mm}{6mm} \hspace {-2mm} Enlarged & 7.3 & 18.3 & 8.0 & 15.6 & 26.3 \\ \hline \hline
\end{tabular}
\label{tab:bgsys}
\end{table*}

\section{Lifetime limit}
\label{calc}
Since no significant event excess was observed in either decay mode, lower limits on the partial lifetime of each have been calculated using a Bayesian method~\cite{limit}.
In this calculation numbers from the ``Enlarged'' rows in Tables~\ref{tab:summary},~\ref{tab:sigsys} and~\ref{tab:bgsys} 
have been used for the signal selection efficiencies, the expected number of background events, and  their systematic uncertainties.
Since the search performance varies depending on the data taking period, separate exposures from \mbox{SK-I} to \mbox{-IV} are taken into account.
The expected number of background events is also different between the two $P_{\rm tot}$ signal regions, and therefore they are considered separately in the following calculation.
The probability density function used for the proton decay rate ($\Gamma$) for the eight signal regions, 
\mbox{SK-I} to \mbox{-IV} each with ``Lower'' and ``Upper'' total momentum regions, is defined as follows:
\begin{equation}
\begin{split}
P_{i}(\Gamma | n_i) = \frac{1}{A_i}\iiint\frac{e^{-(\Gamma \lambda_i \epsilon_i+b_i)}(\Gamma\lambda_i \epsilon_i+b_i)^{n_i}}{n_i!} \\
\times P(\Gamma)P(\lambda_i)P(\epsilon_i)P(b_i)d\epsilon_id\lambda_idb_i,
\end{split}
\end{equation}
where $i$ is the index of each signal region, $A_i$ is a normalization factor representing the total integral of $P_{i}(\Gamma | n_i)$, 
 $n_i$ is the number of observed candidates, $\lambda_i$ is the exposure, $\epsilon_i$ is the signal selection efficiency and $b_i$ is the expected number of background events. 
The prior probability on the decay rate $P(\Gamma)$ is assumed to be uniform
and $P(\lambda_i)$ and $P(\epsilon_i)$ represent the prior probabilities for the exposure and signal selection efficiency, respectively.
Both are assumed to be Gaussian, 
\begin{eqnarray}
P(\lambda_i) \propto 
\begin{cases}
\exp{\left(\frac{-(\lambda_i-\lambda_{0i})^2}{2\sigma^2_{\lambda_i}}\right)}, & (\lambda_i > 0) \\
0, & (\rm{otherwise})
\end{cases} \\
P(\epsilon_i) \propto
\begin{cases} 
\exp{\left(\frac{-(\epsilon_i-\epsilon_{0i})^2}{2\sigma^2_{\epsilon_i}}\right)}, & (\epsilon_i > 0) \\
0, & (\rm{otherwise})
\end{cases}
\end{eqnarray}
where $\lambda_{0i}$ ($\sigma_{\lambda_i}$) and $\epsilon_{0i}$ ($\sigma_{\epsilon_i}$) are the estimates (systematic uncertainties) of the exposure and signal selection efficiency, respectively.
For the expected number of background events the prior probability, $P(b_i)$, is defined as the convolution of a Gaussian and Poisson distribution:
\begin{equation}
P(b_i) \propto 
\begin{cases}
\int_{0}^{\infty}\frac{e^{-B}B^{n_{b_i}}}{n_{b_i}!}\exp{\left(\frac{-(C_{i}b_i-B)^2}{2\sigma^2_{b_i}}\right)} dB, & (b_i > 0) \\
0, & (\rm{otherwise})
\end{cases}
\end{equation}
where $n_{b_i}$ is the expected number of backgrounds in each 500~years atmospheric neutrino MC (before livetime normalization),
$C_i$ is a constant factor to normalize MC to the data livetime and $\sigma_{b_i}$ is the systematic uncertainty on 
the number of background events.
The standard deviation of each Gaussian is taken as the corresponding systematic uncertainty described in Sec.~\ref{results}.
With these definitions the proton decay rate limit at a given confidence level (C.L.) is calculated as
\begin{equation}
{\rm C.L.} = \int^{\Gamma=\Gamma_{{\rm limit}}}_{\Gamma=0}\prod_{i=1}^{8}P_{i}(\Gamma | n_i)d\Gamma.
\end{equation}
As a consequence, lower limits on the partial lifetime are obtained as:
\begin{equation}
\frac{\tau_{{\rm limit}}}{B} = \frac{1}{\Gamma_{{\rm limit}}},
\end{equation}
where $B$ represents the branching ratio of a particular decay mode.
Using the above formulae the resulting limits at 90\% C.L. are $\tau/B(p\rightarrow e^+\pi^0) > 2.4 \times 10^{34}$~years and $\tau/B(p\rightarrow \mu^+\pi^0) > 1.6 \times 10^{34}$~years.
The lifetime limit for the $p\rightarrow e^+\pi^0$ mode has improved by a factor of 1.5 as is expected from the increased exposure.
On the other hand, for the $p\rightarrow \mu^+\pi^0$ mode the limit has improved by a factor of two because one of the two candidates reported in the last paper~\cite{last} moved out of the present analysis's signal region after the updates in the detector calibration described therein.

\section{Conclusion}
\label{conclude}
Improved event selection and reconstruction algorithms have been introduced to enlarge the fiducial volume of the Super-Kamiokande detector from 22.5~kton to 27.2~kton.
With the new selection, the non-neutrino background contamination rate in the additional fiducial volume is kept within 1\%, tolerable level for SK analyses.
The improved search sensitivities for proton decay via $p\rightarrow e^+\pi^0$ and $p\rightarrow \mu^+\pi^0$ are led by the enlarged fiducial volume with the improved event reconstruction algorithm. \par
Using a combined exposure of 450~kton$\cdot$years representing the full Super-Kamiokande data set from \mbox{SK-I} to \mbox{-IV} and data in the additional fiducial volume, we have performed searches for proton decay via $p\rightarrow e^+\pi^0$ and $p\rightarrow \mu^+\pi^0$.
No significant event excess above the atmospheric neutrino backgrounds has been found for either mode.
Lower limits on the partial lifetime of $\tau/B(p\rightarrow e^+\pi^0) > 2.4 \times 10^{34}$~years and $\tau/B(p\rightarrow \mu^+\pi^0) > 1.6 \times 10^{34}$~years are set at 90\% confidence level.
These limits indicate a 1.5 times longer lifetime limit for the $p\rightarrow e^+\pi^0$ mode and two times longer for the $p\rightarrow \mu^+\pi^0$ mode than the previous results~\cite{last}, and are the world's most stringent constraints for these decay modes.

\section*{Acknowledgements}
\input{sk_paper_acknowledgements.tex}


\end{document}

%% file: sk_authors_epi0mupi02019.tex
\newcommand{\AFFicrr}{\affiliation{Kamioka Observatory, Institute for Cosmic Ray Research, University of Tokyo, Kamioka, Gifu 506-1205, Japan}}
\newcommand{\AFFkashiwa}{\affiliation{Research Center for Cosmic Neutrinos, Institute for Cosmic Ray Research, University of Tokyo, Kashiwa, Chiba 277-8582, Japan}}
\newcommand{\AFFipmu}{\affiliation{Kavli Institute for the Physics and
Mathematics of the Universe (WPI), The University of Tokyo Institutes for Advanced Study,
University of Tokyo, Kashiwa, Chiba 277-8583, Japan }}
\newcommand{\AFFmad}{\affiliation{Department of Theoretical Physics, University Autonoma Madrid, 28049 Madrid, Spain}}
\newcommand{\AFFubc}{\affiliation{Department of Physics and Astronomy, University of British Columbia, Vancouver, BC, V6T1Z4, Canada}}
\newcommand{\AFFbu}{\affiliation{Department of Physics, Boston University, Boston, MA 02215, USA}}
\newcommand{\AFFbcit}{\affiliation{Department of Physics, British Columbia Institute of Technology, Burnaby, BC, V5G 3H2, Canada }}
\newcommand{\AFFuci}{\affiliation{Department of Physics and Astronomy, University of California, Irvine, Irvine, CA 92697-4575, USA }}
\newcommand{\AFFcsu}{\affiliation{Department of Physics, California State University, Dominguez Hills, Carson, CA 90747, USA}}
\newcommand{\AFFcnm}{\affiliation{Institute for Universe and Elementary Particles, Chonnam National University, Gwangju 61186, Korea}}
\newcommand{\AFFduke}{\affiliation{Department of Physics, Duke University, Durham NC 27708, USA}}
\newcommand{\AFFfukuoka}{\affiliation{Junior College, Fukuoka Institute of Technology, Fukuoka, Fukuoka 811-0295, Japan}}
\newcommand{\AFFgifu}{\affiliation{Department of Physics, Gifu University, Gifu, Gifu 501-1193, Japan}}
\newcommand{\AFFgist}{\affiliation{GIST College, Gwangju Institute of Science and Technology, Gwangju 500-712, Korea}}
\newcommand{\AFFuh}{\affiliation{Department of Physics and Astronomy, University of Hawaii, Honolulu, HI 96822, USA}}
\newcommand{\AFFicl}{\affiliation{Department of Physics, Imperial College London , London, SW7 2AZ, United Kingdom }}
\newcommand{\AFFkek}{\affiliation{High Energy Accelerator Research Organization (KEK), Tsukuba, Ibaraki 305-0801, Japan }}
\newcommand{\AFFkobe}{\affiliation{Department of Physics, Kobe University, Kobe, Hyogo 657-8501, Japan}}
\newcommand{\AFFkyoto}{\affiliation{Department of Physics, Kyoto University, Kyoto, Kyoto 606-8502, Japan}}
\newcommand{\AFFliv}{\affiliation{Department of Physics, University of Liverpool, Liverpool, L69 7ZE, United Kingdom}}
\newcommand{\AFFmiyagi}{\affiliation{Department of Physics, Miyagi University of Education, Sendai, Miyagi 980-0845, Japan}}
\newcommand{\AFFnagoya}{\affiliation{Institute for Space-Earth Environmental Research, Nagoya University, Nagoya, Aichi 464-8602, Japan}}
\newcommand{\AFFkmi}{\affiliation{Kobayashi-Maskawa Institute for the Origin of Particles and the Universe, Nagoya University, Nagoya, Aichi 464-8602, Japan}}
\newcommand{\AFFpol}{\affiliation{National Centre For Nuclear Research, 02-093 Warsaw, Poland}}
\newcommand{\AFFsuny}{\affiliation{Department of Physics and Astronomy, State University of New York at Stony Brook, NY 11794-3800, USA}}
\newcommand{\AFFokayama}{\affiliation{Department of Physics, Okayama University, Okayama, Okayama 700-8530, Japan }}
\newcommand{\AFFosaka}{\affiliation{Department of Physics, Osaka University, Toyonaka, Osaka 560-0043, Japan}}
\newcommand{\AFFox}{\affiliation{Department of Physics, Oxford University, Oxford, OX1 3PU, United Kingdom}}
\newcommand{\AFFqmul}{\affiliation{School of Physics and Astronomy, Queen Mary University of London, London, E1 4NS, United Kingdom}}
\newcommand{\AFFregina}{\affiliation{Department of Physics, University of Regina, 3737 Wascana Parkway, Regina, SK, S4SOA2, Canada}}
\newcommand{\AFFseoul}{\affiliation{Department of Physics, Seoul National University, Seoul 151-742, Korea}}
\newcommand{\AFFsheff}{\affiliation{Department of Physics and Astronomy, University of Sheffield, S3 7RH, Sheffield, United Kingdom}}
\newcommand{\AFFshizuokasc}{\affiliation{Department of Informatics in
Social Welfare, Shizuoka University of Welfare, Yaizu, Shizuoka, 425-8611, Japan}}
\newcommand{\AFFstfc}{\affiliation{STFC, Rutherford Appleton Laboratory, Harwell Oxford, and Daresbury Laboratory, Warrington, OX11 0QX, United Kingdom}}
\newcommand{\AFFskk}{\affiliation{Department of Physics, Sungkyunkwan University, Suwon 440-746, Korea}}
\newcommand{\AFFtokyo}{\affiliation{The University of Tokyo, Bunkyo, Tokyo 113-0033, Japan }}
\newcommand{\AFFtodai}{\affiliation{Department of Physics, University of Tokyo, Bunkyo, Tokyo 113-0033, Japan }}
\newcommand{\AFFtit}{\affiliation{Department of Physics,Tokyo Institute of Technology, Meguro, Tokyo 152-8551, Japan }}
\newcommand{\AFFtus}{\affiliation{Department of Physics, Faculty of Science and Technology, Tokyo University of Science, Noda, Chiba 278-8510, Japan }}
\newcommand{\AFFtoronto}{\affiliation{Department of Physics, University of Toronto, ON, M5S 1A7, Canada }}
\newcommand{\AFFtriumf}{\affiliation{TRIUMF, 4004 Wesbrook Mall, Vancouver, BC, V6T2A3, Canada }}
\newcommand{\AFFtokai}{\affiliation{Department of Physics, Tokai University, Hiratsuka, Kanagawa 259-1292, Japan}}
\newcommand{\AFFtsinghua}{\affiliation{Department of Engineering Physics, Tsinghua University, Beijing, 100084, China}}
\newcommand{\AFFwu}{\affiliation{Faculty of Physics, University of Warsaw, Warsaw, 02-093, Poland }}
\newcommand{\AFFynu}{\affiliation{Department of Physics, Yokohama National University, Yokohama, Kanagawa, 240-8501, Japan}}
\newcommand{\AFFllr}{\affiliation{Ecole Polytechnique, IN2P3-CNRS, Laboratoire Leprince-Ringuet, F-91120 Palaiseau, France }}
\newcommand{\AFFbari}{\affiliation{ Dipartimento Interuniversitario di Fisica, INFN Sezione di Bari and Universit\`a e Politecnico di Bari, I-70125, Bari, Italy}}
\newcommand{\AFFnapoli}{\affiliation{Dipartimento di Fisica, INFN Sezione di Napoli and Universit\`a di Napoli, I-80126, Napoli, Italy}}
\newcommand{\AFFroma}{\affiliation{INFN Sezione di Roma and Universit\`a di Roma ``La Sapienza'', I-00185, Roma, Italy}}
\newcommand{\AFFpadova}{\affiliation{Dipartimento di Fisica, INFN Sezione di Padova and Universit\`a di Padova, I-35131, Padova, Italy}}
\newcommand{\AFFkeio}{\affiliation{Department of Physics, Keio University, Yokohama, Kanagawa, 223-8522, Japan}}
\newcommand{\AFFwinnipeg}{\affiliation{Department of Physics, University of Winnipeg, MB R3J 3L8, Canada }}
\newcommand{\AFFkcl}{\affiliation{Department of Physics, King's College London, London, WC2R 2LS, UK }}
\newcommand{\AFFwarwick}{\affiliation{Department of Physics, University of Warwick, Coventry, CV4 7AL, UK }}
\newcommand{\AFFral}{\affiliation{Rutherford Appleton Laboratory, Harwell, Oxford, OX11 0QX, UK }}

\AFFicrr
\AFFkashiwa
\AFFmad
\AFFbu
\AFFbcit
\AFFuci
\AFFcsu
\AFFcnm
\AFFduke
\AFFllr
\AFFfukuoka
\AFFgifu
\AFFgist
\AFFuh
\AFFicl
\AFFbari
\AFFnapoli
\AFFpadova
\AFFroma
\AFFkcl
\AFFkeio
\AFFkek
\AFFkobe
\AFFkyoto
\AFFliv
\AFFmiyagi
\AFFnagoya
\AFFkmi
\AFFpol
\AFFsuny
\AFFokayama
\AFFosaka
\AFFox
\AFFral
\AFFseoul
\AFFsheff
\AFFshizuokasc
\AFFstfc
\AFFskk
\AFFtokai
\AFFtokyo
\AFFtodai
\AFFipmu
\AFFtit
\AFFtus
\AFFtoronto
\AFFtriumf
\AFFtsinghua
\AFFwu
\AFFwarwick
\AFFwinnipeg
\AFFynu

\author{A.~Takenaka}
\AFFicrr
\author{K.~Abe}
\AFFicrr
\AFFipmu
\author{C.~Bronner}
\AFFicrr
\author{Y.~Hayato}
\AFFicrr
\AFFipmu
\author{M.~Ikeda}
\author{S.~Imaizumi}
\AFFicrr
\author{H.~Ito}
\AFFicrr 
\author{J.~Kameda}
\AFFicrr
\AFFipmu
\author{Y.~Kataoka}
\AFFicrr
\author{Y.~Kato}
\AFFicrr
% ...


\collaboration{The Super-Kamiokande Collaboration}
\noaffiliation

%% file: sk_paper_acknowledgements.tex


We gratefully acknowledge the cooperation of the Kamioka Mining and Smelting Company.
The Super-Kamiokande experiment has been built and operated from funding by the 
Japanese Ministry of Education, Culture, Sports, Science and Technology, the U.S.
Department of Energy, and the U.S. National Science Foundation. Some of us have been 
supported by funds from the National Research Foundation of Korea NRF-2009-0083526
(KNRC) funded by the Ministry of Science, ICT, and Future Planning and the Ministry of
Education (2018R1D1A3B07050696, 2018R1D1A1B07049158), 
the Japan Society for the Promotion of Science, the National
Natural Science Foundation of China under Grants No. 11235006, the Spanish Ministry of Science, 
Universities and Innovation (grant PGC2018-099388-B-I00), the Natural Sciences and 
Engineering Research Council (NSERC) of Canada, the Scinet and Westgrid consortia of
Compute Canada, the National Science Centre, Poland (2015/18/E/ST2/00758),
the Science and Technology Facilities Council (STFC) and GridPPP, UK, the European Union's 
Horizon 2020 Research and Innovation Programme under the Marie Sklodowska-Curie grant
agreement no.754496, H2020-MSCA-RISE-2018 JENNIFER2 grant agreement no.822070, and 
H2020-MSCA-RISE-2019 SK2HK grant agreement no. 872549.